\documentclass[superscriptaddress,aps,prl,twocolumn, amsmath,amssymb,showpacs]{revtex4-1}
\usepackage{graphicx}  
\usepackage{mathptmx}
\usepackage{epstopdf}
\usepackage{dcolumn}   
\usepackage{bm}        
\usepackage{amssymb}   
\usepackage{amsmath}   
\usepackage{amsthm}
\usepackage{chemarrow}
\usepackage{color}
\usepackage{xcolor} 
\usepackage{mathrsfs}
\usepackage{float}
\usepackage{physics}
\usepackage{bbold}
\usepackage{bbm}
\usepackage{ulem}
\usepackage{latexsym}
\usepackage[a4paper,colorlinks=true,
linkcolor=blue,citecolor=blue,
pdfauthor={ },
pdftitle={ },
pdfsubject={ },
pdfkeywords={ }]{hyperref}

\theoremstyle{plain}
\newtheorem*{theorem*}{Theorem}

\begin{document}


\title{
Constraining Ultralight Dark Matter through an Accelerated Resonant Search
}
\date{\today}

\author{Zitong Xu}
\email[]{These authors contributed equally to this work}
\affiliation{
School of Instrumentation Science and Opto-electronics Engineering, Beihang University, Beijing, 100191, Beijing, China}

\author{Xiaolin Ma}
\email[]{These authors contributed equally to this work}
\affiliation{School of Physics and State Key Laboratory of Nuclear Physics and Technology, Peking University, Beijing, 100871, Beijing, China}

\author{Kai Wei}
\email[Corresponding author: ]{weikai@buaa.edu.cn}
\affiliation{
School of Instrumentation Science and Opto-electronics Engineering, Beihang University, Beijing, 100191, Beijing, China}

\author{Yuxuan He}
\affiliation{School of Physics and State Key Laboratory of Nuclear Physics and Technology, Peking University, Beijing, 100871, Beijing, China}

\author{Xing Heng}
\affiliation{
School of Instrumentation Science and Opto-electronics Engineering, Beihang University, Beijing, 100191, Beijing, China}

\author{Xiaofei Huang}
\affiliation{
School of Instrumentation Science and Opto-electronics Engineering, Beihang University, Beijing, 100191, Beijing, China}

\author{Tengyu Ai}
\affiliation{School of Physics and State Key Laboratory of Nuclear Physics and Technology, Peking University, Beijing, 100871, Beijing, China}

\author{Jian Liao}
\affiliation{School of Physics and State Key Laboratory of Nuclear Physics and Technology, Peking University, Beijing, 100871, Beijing, China}

\author{Wei Ji}
\affiliation{Johannes Gutenberg University, Mainz 55128, Germany}

\author{Jia Liu}
\email[Corresponding author: ]{jialiu@pku.edu.cn}
\affiliation{School of Physics and State Key Laboratory of Nuclear Physics and Technology, Peking University, Beijing, 100871, Beijing, China}
\affiliation{Center for High Energy Physics, Peking University, Beijing, 100871, Beijing, China}

\author{Xiao-Ping Wang}
\affiliation{School of Physics, Beihang University, Beijing, 100191, Beijing, China}
\affiliation{Beijing Key Laboratory of Advanced Nuclear Materials and Physics, Beihang University, Beijing, 100191, Beijing, China}

\author{Dmitry Budker}
\affiliation{Johannes Gutenberg University, Mainz 55128, Germany}
\affiliation{Helmholtz-Institut, GSI Helmholtzzentrum fur Schwerionenforschung, Mainz 55128, Germany}
\affiliation{Department of Physics, University of California at Berkeley, Berkeley, California 94720-7300, USA}

\maketitle

\noindent \textbf{Abstract}

Typical weak signal search experiments rely on resonant effects, where the resonance frequency is scanned over a broad range, resulting in significant time consumption. In this study, we demonstrate an accelerated strategy that surpasses the typical resonance-bandwidth limited scan step without compromising sensitivity. We apply this method to an alkali-noble-gas spin system, achieving an approximately 30-fold increase in scanning step size. Additionally, we obtain an ultrahigh sensitivity of 1.29 ${\rm fT}\cdot{\rm Hz}^{-1/2}$ at around 5 Hz, corresponding to an energy resolution of approximately 1.8$\times10^{-23} {\rm eV}\cdot {\rm Hz}^{-1/2}$, which is among the highest quantum energy resolutions reported. Furthermore, we use this sensor to search for axion-like particles, setting stringent constraints on axion-like particles (ALPs) in the 4.5--15.5 Hz Compton-frequency range coupling to neutrons and protons, improving on previous limits by several-fold. This accelerated strategy has potential applications in other resonant search experiments.

\vspace{4mm}

\noindent \textbf{Introduction}

Precision measurements utilizing resonant effects, such as those found in optical cavities~\cite{opticalcavity}, mechanical resonators~\cite{RevModPhys.91.025005}, and superconducting circuit resonances~\cite{PhysRevLett.132.203601}, are widely employed in detecting weak signals. Ultralight bosonic particles, including axions and axion-like particles (ALPs), are viable dark matter (DM) candidates that exist in galaxies as oscillating fields. These particles can couple with various standard model particles, producing weak signals that motivate both astrophysical and laboratory experiments~\cite{Raffelt:1990yz, Graham:2015ouw, Safronova:2017xyt, Chadha-Day:2021szb, Semertzidis:2021rxs}. 
In our Milky Way galaxy, these particles are typically nonrelativistic with a velocity dispersion of approximately $220 {\rm km}\cdot {\rm s}^{-1}$, leading to a sharp energy spectrum with a spread of about $10^{-6}$~\cite{Kimball2022Bibber}. Consequently, experimental searches for ALPs often rely on resonant effects~\cite{Sikivie:1983ip} to identify a distinct bump-like signal in the noise spectrum. However, when applying these techniques to ultralight bosonic dark matter, the resonance frequency must often be scanned due to the unknown mass of the dark matter. This necessitates tuning the resonant frequency over a wide range, which is both time-consuming and labor-intensive~\cite{Chadha-Day:2021szb}.
Resonant experiments for dark matter detection include those utilizing cavities~\cite{RBF-1987, UF-1990, CAST-2017, HAYSTAC-2017, ADMX-Sidecar-2018, CAPP-2020-1, CAPP-2020-2, ADMX-2021, ADMX-2021-1, HAYSTAC-2021, CAPP-2021, Adair:2022rtw} and nuclear magnetic resonance (NMR) techniques~\cite{Budker:2013hfa, Graham:2013gfa, Wu:2019exd, Garcon:2019inh, Jiang:2021dby, Bloch:2021vnn,abel2017search}. In cavity experiments, the resonance frequency corresponds to the cavity resonant frequency, while in NMR experiments, it corresponds to the Larmor frequency of nuclear spins.

Our current ChangE-NMR experiment employs ultrasensitive atomic magnetometer, integral to precision measurements and the exploration of physics beyond the Standard Model~\cite{Budker:2013hfa, Graham:2013gfa, Wu:2019exd, Garcon:2019inh, Jiang:2021dby, Bloch:2021vnn}. The magnetometer is based on hybrid alkali-noble-gas spin ensembles and operates close to the Spin-Exchange Relaxation-Free (SERF) regime. The advantages of our  alkali-noble-gas spin sensor over other ultrasensitive alkali-only magnetometer and K-$^{3}$He magnetometer are the fine noise suppression and the utilization of $^{21}$Ne spins, whose gyromagnetic ratio is about three orders of magnitude lower than alkali spin and about one order of magnitude lower than $^{3}$He respectively. Typically, comagnetometers such as the self-compensation (SC) comagnetometer \cite{kornack2005nuclear,Wei:2022ggs} and clock-comparison (CC) comagnetometer \cite{Terrano:2021zyh} are used to suppress magnetic noise. However, ultrasensitive SC comagnetometers are limited to a relatively low frequency range due to the low resonance frequency of noble-gas nuclear spin; the sensitivity of CC comagnetometers are not high enough for our ALP searches.  Thus, instead of working in the comagnetometer regime, we use another approach to work in the NMR mode by developing a ferrite magnetic shield with sub-fT magnetic noise and successfully reduced magnetic noise from heating\,\cite{Wei:2022mra}.

\begin{figure*}[htb]
\centering 
    \includegraphics[width=1.5 \columnwidth]{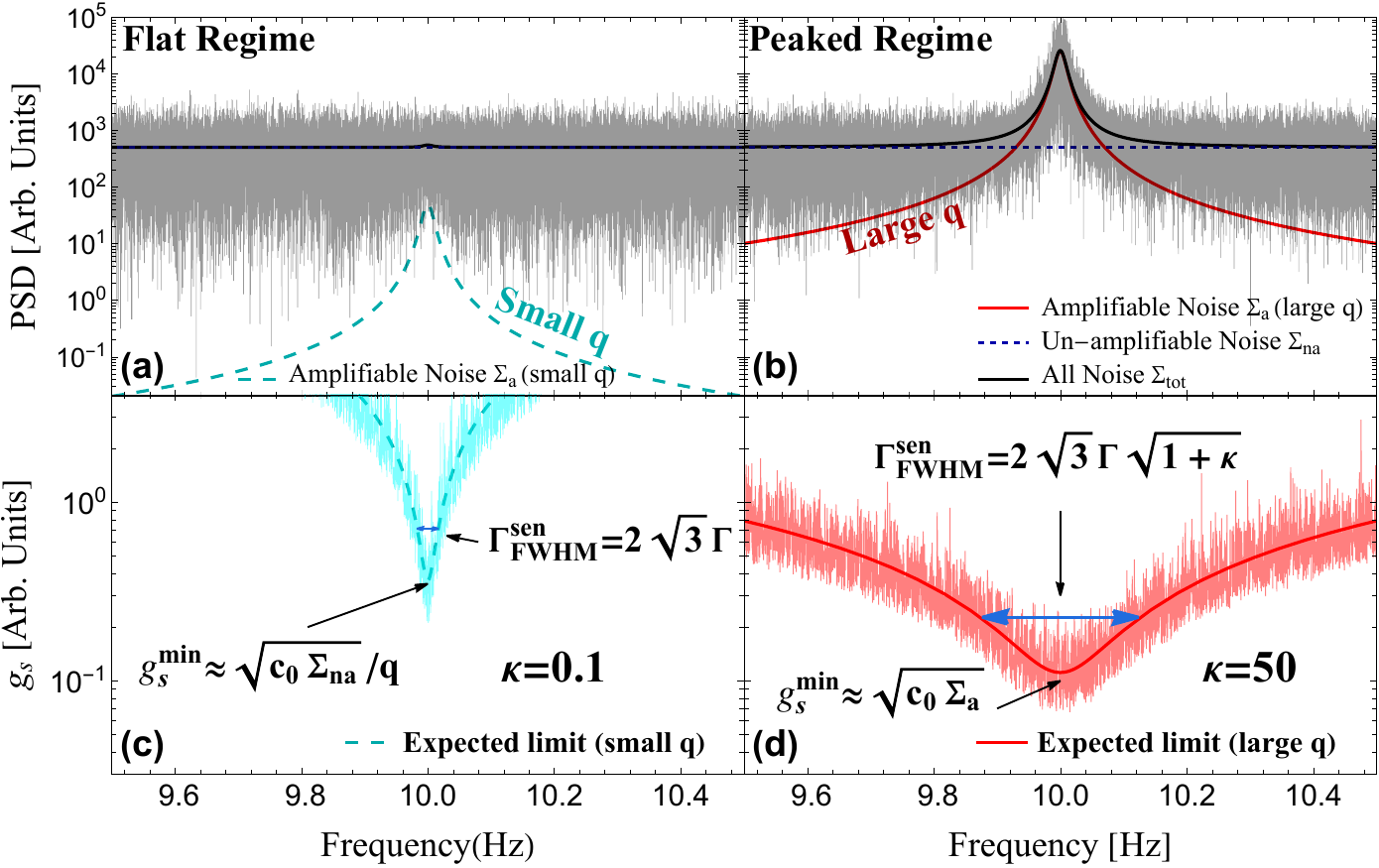} 
\caption{\textbf{Simulated power spectrum density(PSD) and signal sensitivities on $g_{\rm s}$ for the flat and peaked regimes at $f_0 = 10$\,Hz.
} The panels (a) and (b) illustrate the simulated PSD while the panels (c) and (d) show the corresponding signal sensitivities on $g_{\rm s}$ for the flat and peaked regimes at $f_0 = 10$\,Hz, respectively. In both panels, we fix $\Sigma_{\rm a}=0.1$ and $\Sigma_{\rm na}=500$. We set $q=\sqrt{500}$ in flat regime and $q=500$ in peaked regime, leading to different $\kappa $.  The dotted blue lines, dashed lines and black solid lines represent the photon shot noise, magnetic noise after amplification and the total noise. The gray line depicts the simulated total PSD. The bottom panel presents the sensitivity with respect to the signal strength $g_{\rm s}$, showcasing spiky red lines from simulation and a solid thick line from the analytic calculation. 
The full width at half maximum of sensitivity $\Gamma_{\rm FWHM}^{\rm sen}$ and the optimal sensitivity $g_{\rm s}^{\min} $ are illustrated.
	}
	\label{fig:illustration}
\end{figure*}

By sweeping the external magnetic field to scan the resonant frequency, we record a time series of voltage readings from the probe detector. Employing Fourier transformation, we derive the power spectral density (PSD) of the voltage within the frequency domain near the resonant frequency, Fig.\,\ref{fig:illustration}. The PSD typically exhibits a flat profile without the manual injection of fabricated signals~\cite{Jiang:2021dby, Bloch:2021vnn}, as the background noise is predominantly white. Nevertheless, our experiment reveals a distinct peaked PSD. We categorize these two situations as the flat and peaked regimes and provide analytical explanations for their unique characteristics. A similar feature was discussed in Refs.\,\cite{Chaudhuri:2018rqn, Chaudhuri:2019ntz} and used for DM search projections\,\cite{Liu:2018icu, Berlin:2019ahk, Berlin:2020vrk, Brady:2022bus, DMRadio:2022jfv, DMRadio:2022pkf, Brady:2022qne, Benabou:2022qpv}. In the peaked regime, the sensitivity is limited by amplifiable noise, and the high amplification factor does not enhance the sensitivity. However, we discovered that, with the high amplification factor, one can significantly extend the bandwidth while maintaining similar sensitivity levels, between one and two orders of magnitude. We have developed and experimentally verified an expression for this expanded bandwidth. This approach is applicable to various types of resonance experiments, effectively accelerating the scanning process (see Refs.\,\cite{ Dror:2022xpi, Yuzhe2023} for different approaches).

\begin{figure}[tbh]
    \centering
    \includegraphics[width=0.9\columnwidth]{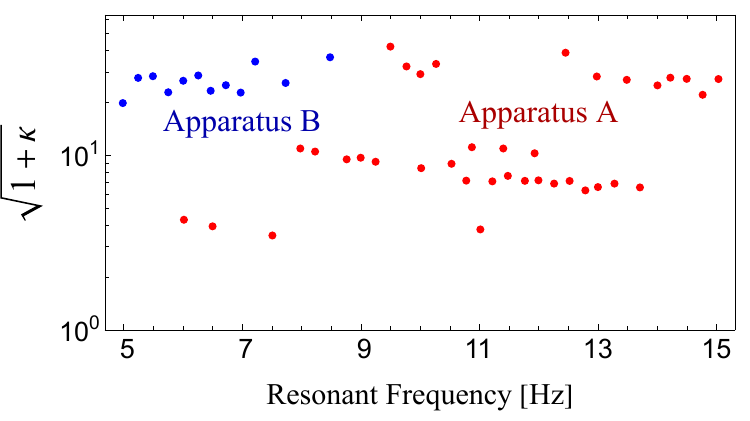}
    \caption{\textbf{
    Scan step boost factor $\left( \sqrt{1+\kappa} \right)$ for 49 individual scans.} The results of Apparatus A is shown in red dots and the results of B are shown in blue.
    }
    \label{fig:acceleration}
\end{figure}

\begin{figure*}[tb]
\centering 
    \includegraphics[width=1.4 \columnwidth]{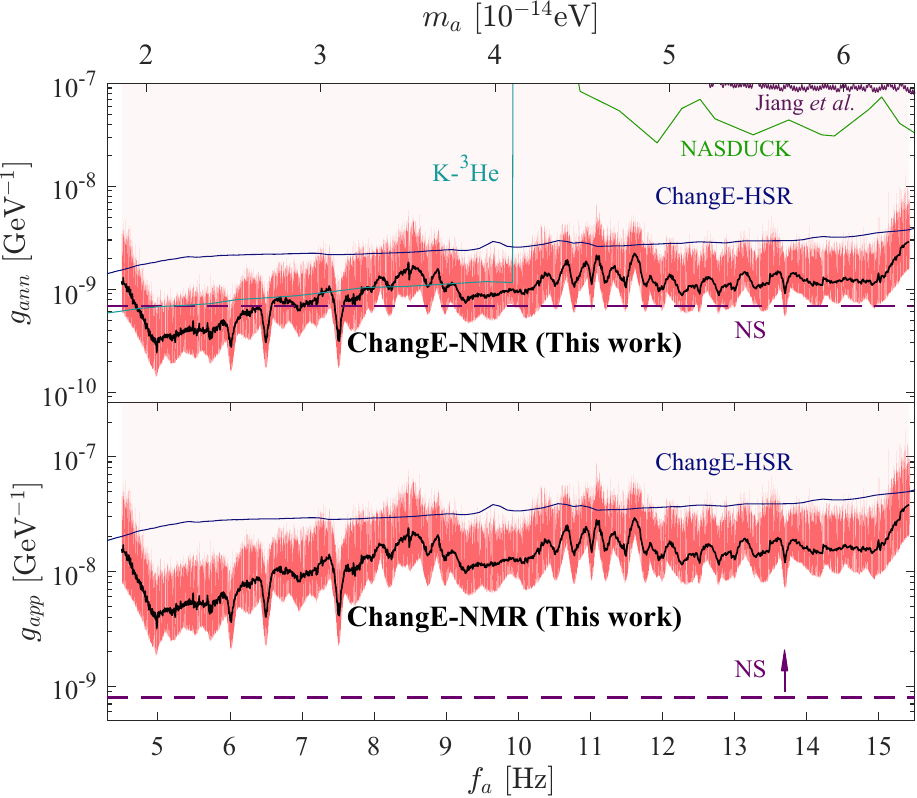}
    \caption{\textbf{The improved limits for neutron coupling $g_{\rm ann}$ and proton coupling $g_{\rm app}$ from the 49 individual scans.} The red spiky line indicates the $95\%$-confidence-level sensitivity for axion-neutron and axion-proton couplings $g_{\rm ann},~g_{\rm app}$, containing approximately 1.5 million test axion-like particle(ALP) frequencies. These frequencies are separated by $\Delta f= 1/(2\tau_a)$, the coherent time $\tau_a \sim 2 \hbar/(m_a v_{\rm vir}^2)$, with $m_a$ and $v_{\rm vir}$ the mass and virial velocity of the ALP. The complete dataset is available at \cite{axiondata}. 
    The black line represents $\pm 250$ bins averaging of the red line to guide the readers' eye. 
    The purple, green, cyan and blue colored lines are previous laboratory searches from Jiang \textit{et al}.~\cite{Jiang:2021dby}, NASDUCK~\cite{Bloch:2021vnn}, K-${}^{3}$He~\cite{Lee:2022vvb}, and ChangE-HSR~\cite{Wei:2023rzs} respectively, 
    while NS is the astrophysics limit from the neutron star cooling~\cite{Buschmann:2021juv}. }
    \label{fig:finalsensitivity}
\end{figure*}

We operate in the NMR mode corresponding to the peaked regime. As a result, this regime results in an $\mathcal{O}(30)$ increase in the full width at half maximum (FWHM) of the sensitivity peak compared to the width of the spin resonance, thus accelerating scan speeds by the same factor, see the scan step boost factor in Fig.~\ref{fig:acceleration}.
We incorporate the stochastic nature of ultralight axion DM into the sensitivity calculation, extending it to the scenario with the magnetometer sensitive to the ALP-field gradient (see below) along two transverse axes.
With a scan step of 0.25\,Hz, much larger than the spin-resonance width of about 0.006\,Hz, our investigations yield improved limits for axion-nucleon couplings within the 5--15\,Hz frequency range, see Fig.\,\ref{fig:finalsensitivity}.

We achieve energy resolution of about 1.8$\times 10^{-23}\, \rm{eV}\cdot{\rm Hz}^{-1/2}$, which is about two orders of magnitude better than the most sensitive alkali-spin magnetometer \cite{kornack2007low}, and about three times better than the most sensitive K-$^3$He comagnetometer \cite{Terrano:2021zyh}.  Exploring sub-eV ($< 10^{17}$ Hz) parameter space for ultralight bosonic dark matter remains meaningful and crucial due to the unknown DM mass, necessitating comprehensive investigation. With our current setup, we achieve optimal sensitivity around $\mathcal{O}(10)$ Hz, offering a valuable window into the parameter space. However, the vastness of this space presents exciting opportunities for future exploration and discovery.
\\

\noindent \textbf{Results}

\noindent \textbf{Properties of Two Regimes.}
Given the sharp spectral peak of the resonant signal, we confine our analysis to a frequency band centered around the peak. Within this band, the noises are flat and can be characterized as random Gaussian white noise. We classify the noise into two categories: non-amplifiable noise $N_{\rm na}$ (e.g., photon shot noise) and amplifiable noise $N_{\rm a}$ (e.g., magnetic noise).

We analyze the signal assuming amplification at the resonant frequency $f_{0}$ by the amplification factor $q$ and Lorentzian amplification function,
\begin{align}
    \eta(f) = \frac{q}{\sqrt{1+ (f-f_{0})^2/\Gamma^2}}\,,
    \label{eq:Lorentzian-amplification}
\end{align}
where $\Gamma$ represents the width of the function.
Due to the independence between noise sources, the total noise variance $\Sigma(f)$ is
\begin{align}
\Sigma(f) = \Sigma_{\rm{na}} + \Sigma_{\rm a} \left[ 1 + \eta(f) \right]^2  + S(f)\,,
\end{align}
where $\Sigma_{\rm{na}}$ and $\Sigma_{\rm{a}}$ denote the variances for non-amplifiable and amplifiable noise, respectively.

The variance of signal takes the form:
\begin{align}
S(f) = g_{\rm s}^2 \times  \eta(f)^2 ,
\end{align}
where $g_{\rm s}$ signifies the signal strength.
For simplicity, we assume the measurement spans a coherence time~\cite{Schive:2014dra,Centers:2019dyn,Gramolin:2021mqv} $\tau_a \sim 2 \hbar/(m_a v_{\rm vir}^2)$, with $m_a$ and $v_{\rm vir}$ the mass and virial velocity of the ALP, ensuring that the signal is confined to a single frequency bin within the data. Consequently, confidence level limits can be established at $95\% $, which yields the expression
\begin{align}
g_{\rm s}^2(f) \lesssim c_0
\left[ \Sigma_{\rm a}
+ \frac{\Sigma_{\rm na}}{q^2} \left( 1+ \frac{(f-f_{0})^2}{\Gamma^2} \right)
\right],
\label{eq:gs-expression}
\end{align}
with $c_0 = 5.99$~\cite{Walck:1996cca},  the experimental premises of $q \gg 1$ and $\Sigma_{\rm na} \gg \Sigma_{ \rm a}$.

For a single measurement at a resonant frequency $f_{0}$, the best sensitivity $g_{\rm s }^{\min}$ is achieved at $f= f_{0}$ and follows a Lorentzian profile. Additionally, we can establish the FWHM $\Gamma^{\rm sen}_{\rm FWHM}$ to characterize the sensitivity bandwidth, requiring $g_{\rm s}(f_{0}+\Gamma^{\rm sen}_{\rm FWHM}/2)=2 g_{\rm s}^{\min}$.
Therefore, we have
\begin{align}
    g_{\rm s}^{\min} = \sqrt{c_0 \Sigma_{a}} \sqrt{1+\kappa^{-1}}, \quad \Gamma_{\rm FWHM}^{\rm sen} \approx 2 \sqrt{3} \Gamma \sqrt{1+ \kappa}\,,
    \label{eq:simple-estimation-of-gmin-Gamma}
\end{align}
with $\kappa \equiv q^2 \Sigma_{\rm a}/\Sigma_{\rm na}$, reflecting the level of amplification.

In Fig.\,\ref{fig:illustration}, we maintain constant noise variances, $\Sigma_{\rm a}$ and $ \Sigma_{\rm na}$, while employing two amplification factors, $q_{1,2}$ with $q_1 \ll q_2$, to illustrate the distinct regimes. Monte Carlo simulations are utilized to generate PSD data for both regimes in the top panel, with corresponding sensitivities presented in the bottom panel. Note that experimentally, the value of $q$ can be controlled via changing the degree of polarization of the noble gas and its coherence time as discussed below.

In the flat regime, $\kappa \ll 1$, the peak sensitivity follows $g_{\rm s}^{\min} \approx \sqrt{c_0 \Sigma_{\rm na}  }/q  $,
which can still be improved by increasing the amplification factor or decreasing the non-amplifiable noise, and it is not affected by the amplifiable noise. The single scan of the experiment can effectively cover the region of $\Gamma_{\rm FWHM}^{\rm sen} \approx 2 \sqrt{3} \Gamma $ which is not improved by increasing $\kappa$. The characteristic of this regime is a flat PSD near the resonant frequency, which is common in the previous cavity experiments (e.g. ADMX~\cite{ADMX:2011hrx, ADMX:2018gho, ADMX:2020ote}, HAYSTAC \cite{Brubaker:2017rna}, CAST-CAPP~\cite{Adair:2022rtw} with flat excess PSD, even for  Superconducting Radiofrequency (SRF) cavity~\cite{Alesini:2022lnp, Cervantes:2022gtv, Tang:2023oid})
and magnetometer experiments (e.g. Jiang \textit{et al}.~\cite{Jiang:2021dby}, NASDUCK~\cite{Bloch:2021vnn}). 
In general, for the flat regime, the scan step is taken to be the largest physical width, the cavity resonant linewidth for cavity, the nuclear linewidth for NMR, and DM linewidth for SRF cavity.

Conversely, in the peaked regime (right panel of Fig.\,\ref{fig:illustration}), $\kappa \gg 1$, the peak sensitivity is approximately $g_{\rm s}^{\min} \approx \sqrt{c_0 \Sigma_{\rm a}  }  $, entirely dictated by the amplifiable noise $\Sigma_{\rm a}$. Reducing  the  non-amplifiable noise $\Sigma_{\rm na}$ and increasing amplification can no longer enhance sensitivity. However, adjusting these parameters can augment the spectral width $\Gamma_{\rm FWHM}^{\rm sen} \approx 2 \sqrt{3} \Gamma \sqrt{\kappa }$ by a factor of $\sqrt{\kappa}$, see also Refs.\,\cite{Chaudhuri:2018rqn, Chaudhuri:2019ntz}.
\vspace{4mm}

\noindent \textbf{ALP Search Principle}.
We seek ALP DM that couples to nuclei through the gradient interaction given by the Hamiltonian
\begin{align}
\mathcal{H} = g_{\rm aNN} \bm{\nabla} a \cdot \bm{\sigma}_N\,,
\label{eq:Hami-nucleus}
\end{align}
where $\bm{\sigma}_N$ represents the nucleus spin, encompassing both neutron and proton contributions, and $g_{\rm aNN}$ signifies the effective axion-nucleus coupling. Consequently, the axion-wind operator induces an anomalous magnetic field $\bm{b}_a \equiv g_{\rm aNN} \bm{\nabla} a /\gamma_\text{N} $, with $\gamma_\text{N} $ denoting the nuclear gyromagnetic ratio.

\begin{figure*}[htb]
\begin{center}
\includegraphics[width=0.95\textwidth]{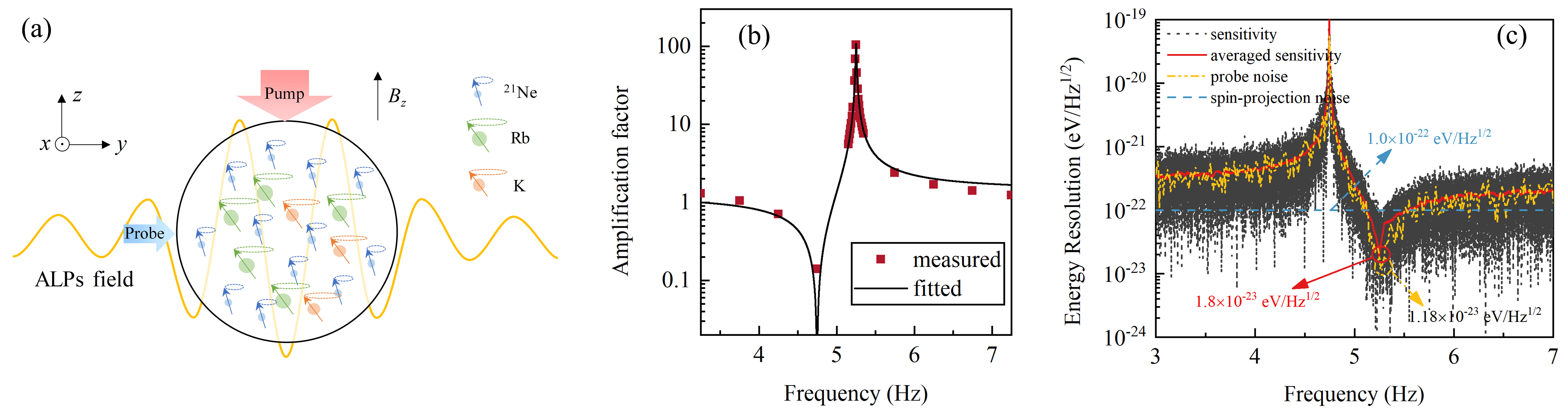}
\caption{\textbf{Principle illustration and measured results for the amplification factor and sensitivity.} Panel (a): The principle of the axion-like particles (ALPs) field search experiment. The ALPs field is experienced by the hybrid spin ensembles and the response to the ALPs field is amplified by the resonance of noble-gas spins and the Fermi-contact enhancement. Panel (b): The amplification factor from Fermi-contact interaction for 1\,pT oscillating \(B_y\) field. The error bars from the standard deviation of three-time measurement are smaller than the data point and are not shown. Panel (c): The sensitivity of magnetic field obtained with Fano lineshape with $f_{0} = 5.25$\,Hz. The black dash curve is the sensitivity curve and the red solid curve is 0.07\,Hz averaged curve, while the yellow dot-dashed line comes from the probe noise.}
\label{fig:FANO-and-amplificationfactor} 
\end{center}
\end{figure*}

We have employed the alkali-noble-gas magnetometer with hybrid spins of $^{21}$Ne-Rb-K which are contained in a glass vapor cell, as shown in Fig.\,\ref{fig:FANO-and-amplificationfactor} (a). Through spin-exchange optical pumping, the pump light polarizes the alkali atoms, which subsequently polarize the noble-gas nuclear spins via spin-exchange interactions with alkali electronic spins. To enhance the polarization homogeneity and efficiency, hybrid optical pumping is employed, following Ref.\,\cite{Babcock:2003zz,romalis2010hybrid}. When we apply a bias field along the pump-light direction, the noble-gas nuclear spins are resonant at the Larmor frequency. The precession of polarized $^{21}$Ne nuclear spins occurs under the influence of ALP fields.  When the ALP-field frequency approaches the resonance frequency, the precession of $^{21}$Ne nuclear spins is maximized. While traditional methods employ pick-up coils or extra magnetometers to read the dipole magnetic field due to the precession of noble-gas nuclear spins, we utilize in-situ alkali spins for this purpose. The Fermi-contact interaction between overlapping alkali and noble-gas spins significantly augments this dipole magnetic field with an amplification factor $q$. 
Therefore, the precession of noble-gas spins due to the ALP field is amplified and read out by in-situ alkali spins. In this Fermi-contact enhanced NMR, the response to the ALP field is amplified by two factors: the resonance of noble-gas spins and the Fermi-contact enhancement. Finally, the precession of alkali spins is measured with linearly polarized probe light via optical rotation.

We employ magnetic fields for calibrating the sensor response to ALP fields. The response to the ALP field, generally different from that to a magnetic field \cite{kornack2002dynamics, ghosh2010measurement, Padniuk:2021dtr}, is determined by relating the responses to magnetic and pseudo-magnetic fields. Due to the pump-light alignment along the $\rm z$ direction, the symmetry of the ${\rm x}$ and ${\rm y}$ components in the Bloch equation causes a 90-degree phase difference between their respective response phases. 
When nuclear spin precession frequency significantly exceeds its relaxation rate, the nuclear spin projection shifts between \textit{x} and \textit{y} axis rapidly, making the projections nearly equal, see Methods\,\cite{deSalas:2020hbh, Cowan:2010js, Brown:2016ipd, Almasi:2018cob}.

However, the lineshape of amplification for usual magnetic field is not entirely symmetric due to the well-known Fano resonance \cite{miroshnichenko2010fano, Jiang:2023nan}, as illustrated in Fig.\,\ref{fig:FANO-and-amplificationfactor} (b). When both resonant ($^{21}$Ne) and continuous-background (alkalis) transitions coexist in a single system, their interplay results in Fano resonances.  The amplification function, incorporating the Fano effect, is described by
\begin{align}
\eta_{\rm F}(f)=\sqrt{\frac{\Gamma^2+(q\Gamma+f_0-f)^2}{\Gamma^2+(f-f_0)^2}}\,
\label{eq:Fano-amplification}
\end{align}
with the resonant frequency $f_0$ and the relaxation rate $R_2^{\rm n}=(T_\text{2n})^{-1}=2\pi \Gamma$. 
The amplification factor $q$ is primarily determined by the nuclear spin coherence time $T_\text{2n}$ and the effective magnetic field $B_0^{\rm n}$, represented as $q=\frac{1}{2} \gamma_\text{Ne} B_0^{\rm n} T_\text{2n}$ in Methods. The magnitude of $q$ determines the degree of asymmetry, with a higher $q$ indicating a more pronounced resonance. As $q$ approaches infinity, the Fano lineshape transforms into a symmetric Lorentz lineshape in Eq.\,\eqref{eq:Lorentzian-amplification} as $\eta_{\rm F}(f) \to 1 + \eta(f)$.

In Fig.\,\ref{fig:FANO-and-amplificationfactor} (b), we set the resonance frequency to $4.98$\,Hz and measure the  $q$. We fit the measured $q$ with Eq.\,\eqref{eq:Fano-amplification} and find it to be in good agreement with the experimental parameters $B_0^{\rm n} \simeq 250$\,nT and $T_{\rm 2n} \simeq 31$\,s. 
In Fig.\,\ref{fig:FANO-and-amplificationfactor} (c), we show the  noise spectrum with a Fano resonance seen. The Fano resonance has minimal impact on sensitivity at the resonance frequency, but significantly affects the non-resonance region.  For a usual magnetic field, Fano lineshape accurately describes the response of the system. However, since we only consider the ALP field coupling to the nuclear spins, the response is more accurately described by a Lorentzian shape. A detailed analytical derivation and simulations for both usual magnetic fields and ALP fields are provided in Methods.

The measured sensitivity is shown in Fig.\,\ref{fig:FANO-and-amplificationfactor} (c).  The energy resolution at 5.25\,Hz reaches $1.8 \times 10^{-23}\,\text{eV}\cdot{\rm Hz}^{-1/2}$. The best sensitivity is achieved at the gain peak seen in Fig.\,\ref{fig:FANO-and-amplificationfactor} (b), where 
the amplification exceeds 100-fold. Here, the sensitivity is limited by the (amplified) magnetic noise. However, the Fano effect arising from the interference between resonant ($^{21}$Ne) and continuous-background (alkali) transitions, renders a narrow region on the left side of the resonant peak insensitive to signals (see also \cite{Jiang:2023nan}). Analyzing the averaged sensitivity curve reveals that beyond the resonant region, the averaged sensitivity hovers around $\sim 1 \times 10^{-22}\,\text{eV}\cdot{\rm Hz}^{-1/2}$, influenced by non-amplifiable noise, probe background noise (shown with dot-dashed line), which is close to the theoretical value of photon shot noise. The sensitivity achieved is below the spin projection noise and photon shot noise of the alkali metal magnetometer utilizing Fermi contact amplification \cite{Budker2008PRA}.

\vspace{4mm}

\noindent \textbf{ALP Search Results}.
We conducted 49 individual scans within a step interval of approximately 0.25\,Hz, chosen according to Eq.\,\eqref{eq:simple-estimation-of-gmin-Gamma}, with a cumulative measurement duration of 1440 hours, over the frequency range of 5--15\,Hz. Employing frequency-domain analysis outlined in Ref.\,\cite{Lee:2022vvb}, we searched for ALP signals taking in consideration the stochastic effects \cite{Centers:2019dyn}. It is important to note a distinction from Ref.\,\cite{Lee:2022vvb}: while their investigation involves a sole sensitive axis, our study incorporates both the $x$ and $y$ axes as sensitive directions in the NMR mode. Consequently, we have adapted the algorithm accordingly.

When compared to the previous simple estimation, the Fano effect can significantly modify the response of the system to magnetic fields, while having less impact on the response to ALP. In Methods, we account for this effect and calculate the sensitivity bandwidth and the best-sensitivity frequency,
 \begin{align}
 & \Gamma_{\rm FWHM} =2\sqrt{3}\Gamma 
     \sqrt{1+\kappa \left(1 - 
     \frac{\kappa (\kappa + 2 q^2)}{(\kappa + q^2)^2} \right) } \, , 
          \label{eq:Fano-FWHM} \\
 & \tilde{f}_0  = f_0 + \Gamma \frac{\kappa q}{\kappa + q^2}\,   .  
     \label{eq:Fano-f0}
 \end{align}
Because $\Sigma_{\rm na} \gg \Sigma_{\rm a}$ we have $q^2 \gg \kappa$, thus Eq.\,\eqref{eq:Fano-FWHM} reverts to Eq.\,\eqref{eq:simple-estimation-of-gmin-Gamma}.

\begin{figure}[htb]
 \includegraphics[width=0.99\columnwidth]{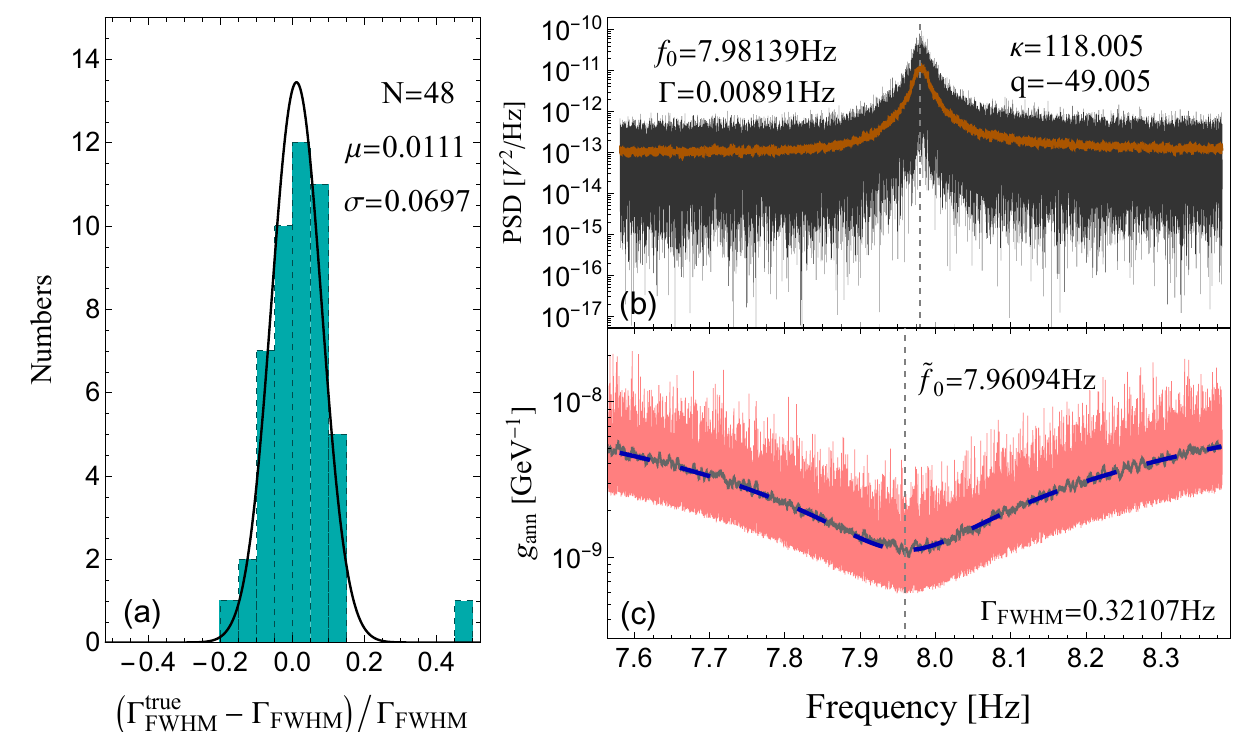} 
    \caption{\textbf{The matching between the model and the experiment.}
    Panel (a): Histogram of  deviation between the analytic estimate and the experimental value for $\Gamma_{\rm FWHM}$. A Gaussian fit indicates a difference between theory and experiment within $7\%$. A single measurement with poor quality is omitted from the fitting process and sensitivity calculation.
    Panels (b) and (c): The power spectrum density(PSD) and sensitivity of the data from a resonance scan at $f_0 =7.98139$\,Hz are shown as black and red spiky lines, together with brown and gray lines for their average. The blue dashed line shows the theory prediction for $g_{\rm ann}$.}
    \label{fig:FWHM-width-compare}
\end{figure}

In the experiment, we have two apparatus, denoted as A and B in Methods, collecting the data. Apparatus A has 
$\left|q\right| \sim 40-50$ and $\kappa$ in the range of 10--1000, while apparatus B has $\left|q\right| \sim 100$ and $\kappa$ ranging between 400 and 1000. We obtain the values of $q,~\Gamma$ from the calibration process and extract $\kappa$ from the data. Therefore, we can compute $\Gamma_{\rm FWHM}$ and $\tilde{f}_0$ and compare them with the experiment results.
In Fig.\,\ref{fig:FWHM-width-compare}, we observe that the theory and experimental results are in good agreement for $\Gamma_{\rm FWHM}$, differing by only $7\%$. Meanwhile, the actual PSD and sensitivity are shown, confirming the shift of $\tilde{f}_0$ and the enhancement of $\Gamma_{\rm FWHM}$.

After computing the sensitivity, we evaluate the ALP-field nucleon couplings as illustrated in Fig.\,\ref{fig:finalsensitivity}. The coupling constants $g_{\rm ann}$ for neutrons and $g_{\rm app}$ for protons are derived following \cite{Wei:2023rzs}. The sensitivity of every scan spans over a range of a few tenth of a hertz. 
Our sensitivity surpasses our prior ChangE-HSR study by a few-fold in both couplings. Additionally, our results exceed the astrophysical bounds within the frequency range of 4.6--6.6\,Hz and around 7.5\,Hz.

The goal of our investigation is detection of possible ALP-dark-matter signals in our dataset. We undertake a post-analysis to evaluate the significance of the best-fit signal in comparison to the background-only model and to statistically assess data points surpassing the $5\sigma$ threshold~\cite{Lee:2022vvb, Wei:2023rzs}. 
We found several candidates surpass the $5\sigma$ threshold of the background-only hypothesis, with consideration for the look-elsewhere effect. However, each of these candidate frequencies can be tested by multiple adjacent individual scans. Notably, these $5\sigma$ peaks were not consistently present in other individual scans, implying potential transient background fluctuations. Consequently, we conclude the absence of ALP DM signals in our experiments.

In summary, we conducted a search for ultralight ALP DM using a magnetometer operating in the Fermi-contact enhanced NMR mode. Our findings revealed that the experiment operates in the ``peaked'' regime, dominated by magnetic noise. In this regime, we optimize the scanning speed by choosing a step size exceeding the intrinsic NMR linewidth by approximately a factor of 30, as determined by the amplification parameter and the noise level in the system. We analytically elucidated the properties of this peaked regime and found that the sensitivity bandwidth inferred from the parameters of the system agrees with experimental results at a $7\%$ level. Furthermore, we established stringent constraints on the coupling of ALP to both neutrons and protons, surpassing previous searches by a few-fold. Our laboratory results surpass the astrophysical ones derived from neutron-star cooling in the frequency range of 4.6--6.6\,Hz and around 7.5\,Hz. 

\vspace{4mm}

\noindent \textbf{Discussion}

This search is rooted in the NMR regime, which diverges significantly from the previous self-compensation (SC) regime \cite{wei2022constraints, wei2022ultrasensitive} and the hybrid-spin-resonance (HSR) regime \cite{Wei:2023rzs}, despite their common basis in the K-Rb-$^{21}$Ne spin ensembles. In the SC regime, the magnetization of noble-gas nuclear spins compensates for low-frequency magnetic noise under a certain external bias field, where the value equals the sum of the Fermi-contact interaction fields of alkali spins and noble-gas nuclear spins. There is no physical spin resonance in the SC regime. In contrast, in the HSR regime, the alkali spins and noble-gas nuclear spins are strongly coupled by setting the bias field to a specific value, which decreases the total magnetic field experienced by the alkali spins, distinguishing it from the SC regime. Actually In the HSR regime, the dynamic performance of noble-gas nuclear spins is significantly amplified by several orders of magnitude through coupling to the highly dynamic alkali spins, resulting in resonance between noble-gas nuclear spins and alkali spins, where their Zeeman splittings are similar to each other.
However, in the NMR regime, the bias magnetic field is tuned such that the Zeeman splitting of the nuclear spin matches the dark matter mass, enabling resonant absorption of the dark matter particles. To explore a wide range of dark matter masses, the bias field has to be scanned. In this experiment, the resonance frequencies of noble-gas nuclear spins and alkali spins are widely separated by several kHz, unlike the HSR regime. Moreover, there is no compensation for magnetic noise compared to the SC regime. Furthermore, in the NMR regime, the resonance is asymmetric, with a profile significantly different from the symmetric Lorentzian shape, attributed to the well-known Fano resonance. Therefore, the NMR regime search differs from searches conducted under the SC and HSR regimes.

In resonant dark matter search experiments, achieving both high sensitivity and broad coverage is often challenging. While maintaining world-leading sensitivity can be difficult, achieving a wide search range is also not straightforward. In cases where the measurement sensitivity is limited by amplifiable noise, increasing the response amplitude (usually denoted as the amplification factor $q$) does not necessarily improve the sensitivity. However, we found that increasing the amplification factor can extend the measurement bandwidth. We formulated expressions for the expanded bandwidths, determined by the amplification factor and the ratio of amplifiable noise to non-amplifiable noise. The validity of the bandwidth-expansion method was verified using ultrasensitive hybrid spin-based sensors. This approach can be applied also to other types of resonance experiments to broaden the bandwidth. The analysis will guide the design of future, higher-sensitivity experiments.

The ADMX and CAPP experiments have achieved unprecedented sensitivity in probing the Quantum chromodynamics(QCD) axion region through their investigations of the axion-photon coupling~\cite{ADMX:2021nhd,Yang:2023yry}. Similarly, the JEDI Collaboration introduced a storage ring for pioneering axion dark matter research~\cite{JEDI:2022hxa}. These experiments may target vastly disparate frequencies, resulting in varying coherence times for ALP dark matter detection. For instance, our experiment targets frequencies around 10 Hz, whereas cavity experiments often operate near 1 GHz. This discrepancy leads to eight orders of magnitude differences in the coherent time.
Additionally, the scan step is influenced by experiment quality factors. For experiments with quality factors lower than the dark matter's \( \sim 10^6 \), the scan step typically equals the measurement width of the system. Conversely, experiments with higher quality factors (e.g., superconducting cavities) often adopt scan steps equivalent to the dark matter signal width. This difference affects the number of data points measured and, in conjunction with total measurement time, dictates search coverage.

\vspace{4mm}

\noindent \textbf{Methods}

In Methods section, we begin by providing a detailed description of the experimental setup and its fundamental underlying principles. This includes a discussion on the manifestation of the Fano effect within our experiment and its impact on our results. Subsequently, we delve into the stochastic nature of ultralight axion DM and its implications. Finally, we explore the statistical methods employed to calculate signal sensitivity and outline the post-analysis process for discriminating potential axion DM candidates. 

\vspace{4mm}

\begin{figure}[htb]
    \centering
    \includegraphics[width=3.4 in]{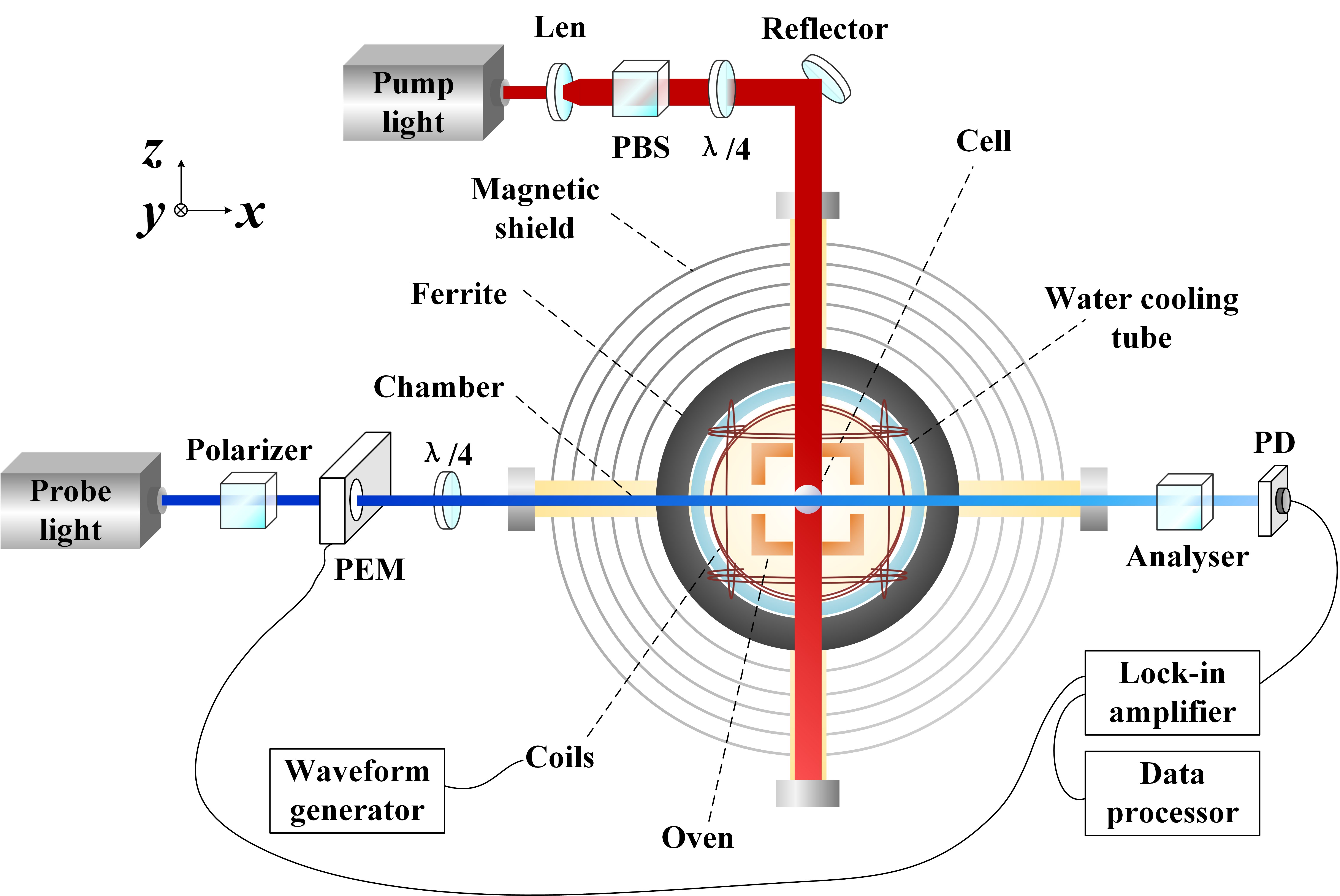}
    \caption{\textbf{Experimental setup.} PBS, polarizing beam splitter; PD, photo-diode detector; PEM, photoelastic modulator.}
    \label{fig:experimental setup}
\end{figure}

\noindent \textbf{Experiment Setup}.
The experimental setup for the axion DM search is illustrated in Fig.\,\ref{fig:experimental setup}. A spherical vapor cell with a diameter of 12 mm containing alkali-metal atoms and noble-gas atoms is at the center of the apparatus. The cell is heated to 468\,K with a several flexible heating films driven by AC current. The films are arranged to compensate the magnetic field produced by the heating current. The cell and oven are inside a polyetheretherketone (PEEK)  chamber, and a water-cooling tube is wrapped around the PEEK chamber to prevent the ferrite outside the chamber from heating up and generating thermo-magnetic noise. The geomagnetic environment around the cell is shielded by a five-layer magnetic shield and magnetic compensation coils wrapped on the PEEK chamber. Atomic spins are polarized with circularly polarized pump light tuned to the D1 line of K atoms propagating along the \textit{z} axis. Alkali-atom polarization is transferred to the noble-gas nuclei in spin-exchange collisions. The precession information is read out with a linearly polarized probe light propagating along \textit{x}, which is detuned towards longer wavelengths by 0.5\,nm from the D1 line of Rb atoms. In the presence of a signal, which is rotation of the nuclear spins, the polarization direction of the linearly polarized probe light is rotated upon passing through the cell, and the optical rotation angle can be read by detecting the transmitted light intensity. The probe light polarization is modulated with a photoelastic modulator (PEM) at a frequency of 50\,kHz to suppress low-frequency noise. The transmitted intensity detected with a photodiode is demodulated with a lock-in amplifier.  In order to suppress low-frequency interference caused by air convection, the cell, the magnetic shield system and the optical system are placed in a water-cooling box (not shown in the figure).

\vspace{4mm}

\noindent \textbf{The Responses to Magnetic Field and ALP Field}.
The spin dynamics of the Fermi-contact enhanced spin resonator  could be described by the coupled Bloch equations \cite{kornack2005nuclear, Padniuk:2021dtr,PhysRevLett.121.261803}:

\begin{eqnarray}\label{eq.Bloch}
   \frac{{\partial  {\bf{P}^{\rm e}} }}{{\partial t}} &=& \frac{{ \gamma _{\rm e} }}{Q}\left({\bf{B}} + \lambda  {{M}}^{\rm n}_0  {\bf{P}}^{\rm n} + {\bf{\beta^{e}}} +{\Omega}\frac{Q }{\gamma _{\rm e} } \right) \times  {\bf{P}}^{\rm e}  \nonumber\\
   &&+ \frac{(P^{\rm e}_\text{z0}\hat{z} -  {\bf{P^e}} ) \{ R_1^{\rm e}, R_2^{\rm e} ,R_2^{\rm e} \}}{Q},\nonumber \\
   \frac{{\partial  {\bf{P^{\rm n}}} }}{\partial { t}} &=&  \gamma _\text{Ne} \left({\bf{B}} + \lambda  M^{\rm e}_0  {\bf{P}}^{\rm e} + {\bf{\beta^\text{Ne}}} + \frac{\Omega }{\gamma _\text{Ne}} \right) \times {\bf{P}}^{\bf{n}}  \nonumber\\
   &&+ (P^{\rm n}_\text{z0}\hat{z} -{\bf{P^{\rm n}}} )\{  R_1^{\rm n} , R_2^{\rm n} ,R_2^{\rm n} \}\,,
\end{eqnarray}
where \(\gamma _{\rm e} \) and \(\gamma _\text{Ne}\) are gyromagnetic ratios of electrons and $^{21}$Ne nucleons, respectively. $Q$ is the slowing-down factor arising from spin-exchange collisions and hyperfine interaction, $\bf{P^e}$ and $\bf{P^{\rm n}}$ are the spin polarizations of alkali electron  and  $^{21}$Ne nucleons, respectively. 
$\bf{B}$ and $\Omega$  are external magnetic field and inertial rotation. $\bf{\beta}^{\rm e}$ and $\bf{\beta}^{\rm Ne}$ are ALP-fields couplings to alkali electron spins and noble-gas nuclear spins, respectively. Here we focus on the $\bf{\beta}^{\rm Ne}$. The Fermi-contact interaction between alkali atoms and $^{21}$Ne atoms can be described by an effective magnetic field $\lambda M^{\rm e,n}_0  {\bf{P}}^{\rm e,n}$, where $M^{\rm n}_0$ ($M^{\rm e}_0$) is the maximum magnetization of $^{21}$Ne nucleon (alkali electron). For a uniformly spin polarized spherical cell, $\lambda = 8\pi\kappa_0/3$, where $\kappa_0$ is the enhancement factor. $P_{\rm z0}^{\rm e}$ and $P_{\rm z0}^{\rm n}$ are steady electronic and nuclear spin polarization. $R_1^{\rm e}$ and $ R_2^{\rm e}$ are the longitudinal and transverse relaxation rates for alkali electron spin, respectively, and $R_1^{\rm n}$ and $R_2^{\rm n}$ are the longitudinal and transverse relaxation rates for the $^{21}$Ne nucleon spin. 

The response of the nuclear spin to the ALP field in the rotating coordinate system can be obtained as

\begin{align}
\tilde P_ \bot ^{\rm n} = \frac{{{\rm i}\mathop \gamma_\text{Ne} P_{\rm z}^{\rm n} \tilde\beta_ \bot ^\text{Ne}}}{{ {{\rm i}\left( {{\omega _{\rm n}} - {\omega}} \right) - R_2^{\rm n}} }}\,.
\end{align}
Here, $\tilde P_ \bot ^{\rm n}=\tilde P_{\rm  x} ^{\rm n}+{\rm i} \tilde P_ {\rm y} ^{\rm n}$ , $\tilde\beta_ \bot ^\text{Ne}=\tilde\beta_{\rm x} ^\text{Ne}+{\rm i}\tilde\beta_{\rm y} ^\text{Ne}$, \({\omega _{\rm n}} = {\gamma _\text{Ne}}\left( {{B_{\rm z}} + \lambda M_0^{\rm e}P_{\rm z0}^{\rm e}} \right)\) is the Larmor frequency of nuclear spin, and $\omega$ is the frequency of the ALP field. $R_2^{\rm n}=(T_\text{2n})^{-1}$. Then the effective nuclear magnetic field arising in response to the ALP field can be expressed as $\tilde B_\bot^{\text{eff}}=\lambda M_0^{\rm n} \tilde P_ \bot ^{\rm n}$. Thus, the response of electronic spin induced by the nuclear response magnetic field can be obtained as
\begin{align}
\tilde P_ \bot ^{\rm e} &= \frac{{{\rm i}\frac{{\mathop \gamma \nolimits_{\rm e} }}{Q}P_{\rm z}^{\rm e} \tilde B_\bot^{\text{eff}}}}{{ {{\rm i}\left( {{\omega _{\rm e}} - {\omega}} \right) - \frac{{R_2^{\rm e}}}{Q}}}} \,, \nonumber\\
&=\frac{{ - \mathop \gamma _\text{Ne} \frac{{\mathop \gamma \nolimits_{\rm e} }}{Q}P_{\rm z}^{\rm e} B_{\rm z}^{\rm n} \tilde\beta_ \bot ^\text{Ne}}}{{\left[ {{\rm i}\left( {{\omega _{\rm n}} - {\omega }} \right) - R_2^{\rm n}} \right]\left[ {{\rm i}\left( {{\omega _{\rm e}} - {\omega }} \right) - \frac{{R_2^{\rm e}}}{Q}} \right]}}\,,
\label{eq:pi/2 argue}
\end{align}
with \({\omega _{\rm e}} = \frac{{{\gamma _{\rm e}}}}{Q}\left( {{B_{\rm z}} + \lambda M_0^{\rm n}P_{\rm z0}^{\rm n}} \right)\). It can be seen that the coefficients of $\beta_ {\rm x} ^\text{Ne}$ and $\beta_{\rm y} ^\text{Ne}$ only differ by one imaginary unit, so the \textit{x} and \textit{y} axes are symmetric, and the responses of nuclear and electronic spins to the ALP fields in the \textit{x} and \textit{y} directions differ by 90 degrees of phase.

The ALP field is equivalent to an oscillating magnetic field which couples to nuclear spins, so the possibility of using the oscillating magnetic field to calibrate the ALP field is analyzed. Magnetic or non-magnetic perturbations can be expressed in a complex form. Based on the symmetry of the \textit{x} and \textit{y} axes, the transverse input can be expressed as \(B_\perp = B_{\rm x} + {\rm i} B_{\rm y}\) and  \(\beta_\perp ^\text{Ne} = \beta_{\rm x}^\text{Ne} + {\rm i} \beta_{\rm y} ^\text{Ne}\), respectively. Thus, the response of the sensor corresponding to the two parts of the signal can be calculated separately as \(M_ + ^{\rm e}\mathop {\rm e}\nolimits^{{\rm i}\omega t}\)  and \( M_ - ^{\rm e}\mathop {\rm e}\nolimits^{ - {\rm i}\omega t} \), where

\begin{align}
M_ + ^{\rm e} &=&& \frac{{\frac{{{\gamma _{\rm e}}}}{Q}P_{\rm z}^{\rm e}\left[ {\left( {\omega  - {\omega _{\rm n}} - {\rm i}R_2^{\rm n} - {\omega _\text{nn}}} \right){B_ \bot } - {\omega _\text{nn}}\beta_\perp ^\text{Ne}} \right]}}{{\left( {{\rm i}\omega  - {\rm i}{\omega _{\rm e}} + \frac{{R_2^{\rm e}}}{Q}} \right)\left( {{\rm i}\omega  - {\rm i}{\omega _{\rm n}} + R_2^{\rm n}} \right) + {\omega _\text{en}}{\omega _\text{ne}}}}, \notag\\
M_ - ^{\rm e} &=&& \frac{{\frac{{{\gamma _{\rm e}}}}{Q}P_{\rm z}^{\rm e}\left[ {\left( { - \omega  - {\omega _{\rm n}} - iR_2^{\rm n} - {\omega _\text{nn}}} \right){B_ \bot } - {\omega _\text{nn}}\beta_\perp ^\text{Ne}} \right]}}{{\left( { - {\rm i}\omega  - {\rm i}{\omega _{\rm e}} + \frac{{R_2^{\rm e}}}{Q}} \right)\left( { - {\rm i}\omega  - {\rm i}{\omega _{\rm n}} + R_2^{\rm n}} \right) + {\omega _\text{en}}{\omega _\text{ne}}}}\,.
\end{align}

\begin{figure}[htb]
    \centering
    \includegraphics[width=0.98 \columnwidth]{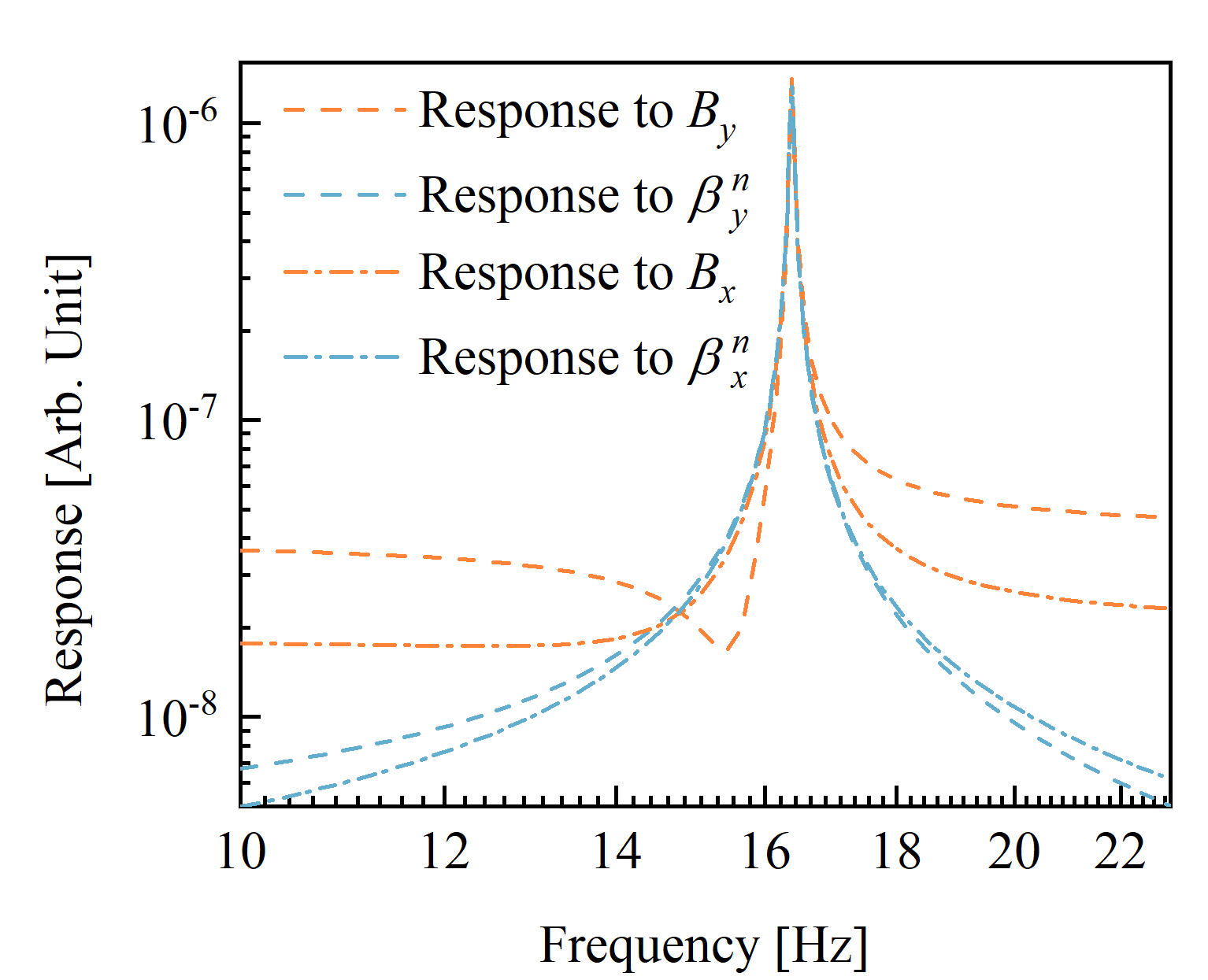}
    \caption{\textbf{Responses to usual magnetic field and axion-like particles (ALPs) field with a bias magnetic field of 5000\,nT.} The responses at the resonance peak to both usual magnetic field and ALPs field are equal.}
    \label{fig:response to B and bn}
\end{figure}

\noindent Here, \({\omega _\text{{en}}} = \frac{{{\gamma _{\rm e}}}}{Q}\lambda M_0^{\rm n}P_\text{z0}^{\rm e}, {\omega _\text{{ne}}} = {\gamma_\text{Ne}}\lambda M_0^{\rm e}P_\text{z0}^{\rm n}, \omega_\text{nn}={\gamma_\text{Ne}}\lambda M_0^{\rm n} P_\text{z0}^{\rm n}\). When the ALP field or the magnetic field is resonant with the nuclear spins, the coefficients of responses to ALP field and magnetic field are approximately equal (when $R_2^{\rm n}\ll \omega_\text{nn}$). Therefore, the ALP field can be calibrated with an oscillating magnetic field. Since the sensor detects the information of the \textit{x}-axis electron spin polarization, the \textit{x}-direction information can eventually be read by separating the real part as \(P_{\rm x}^{\rm e} = {\mathop{\rm Re}\nolimits} \left( {M_ + ^{\rm e}{{\rm e}^{{\rm i}\omega t}} + M_ - ^{\rm e}{{\rm e}^{ - {\rm i}\omega t}}} \right)\). According to the solution, the response to usual magnetic field and ALP field can be simulated as shown in Fig.\,\ref{fig:response to B and bn}. It can be seen that the nuclear spin resonance frequency is 16.38\,Hz when a 5000\,nT bias magnetic field is applied, and the response of the nuclear spin to the usual magnetic field as well as the ALP field is almost equal at the resonance. It is noteworthy that the response to the usual magnetic field is asymmetric because both nuclear and electronic spins respond to the usual magnetic field \cite{Jiang:2023nan}. However, the response to ALP field is more symmetric because there is only discrete resonance (due to $^{21}$Ne) in the system. Thus, the response to usual magnetic field is fitted by Fano lineshape and the response to the ALP field is calibrated by the Lorentz lineshape in the small range around the resonance peak.

\vspace{4mm}

\begin{figure}
 \centering
    \includegraphics[width=3.4in]{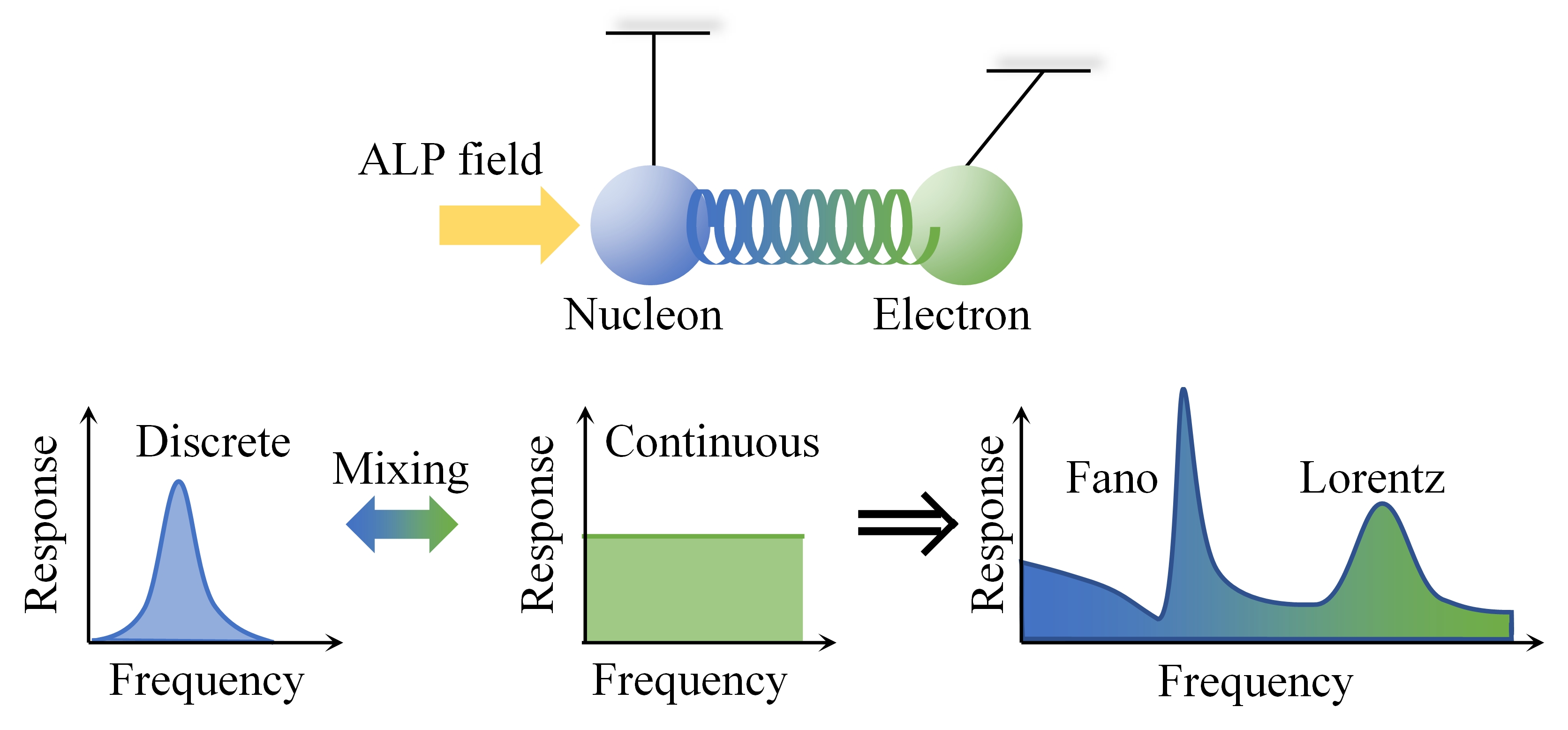}
    \caption{\textbf{The illustration of how the Fano effect modifies the spectrum of the nuclear spin.} The Fermi-contact interaction between the nuclear and electronic spins are like a spring. The responses of the spin ensembles mix and show a Fano effect.}
    \label{fig:fano principle}
\end{figure}

\noindent \textbf{FANO Effect and Its Response}.
The magnetic field response can be used to calibrate the ALP field. In practice we find that the exact lineshape of usual magnetic field amplification as a function of the test frequency is asymmetric with the profile significantly distorted compared to the symmetric Lorentzian shape, consistent with the simulation in Fig.\,\ref{fig:response to B and bn}. This asymmetric lineshape is caused by the well-known Fano resonance. Specifically, \text{Rb} and $^{21}\text{Ne}$ spins both experience the measured oscillating usual magnetic field and both generate signatures at the frequency of the measured field, and the asymmetric shape is caused by the interference of the response of the two spin ensembles as shown in Fig.\,\ref{fig:fano principle}. The resonance frequencies of noble-gas and alkali-metal atoms are different, similar to pendulums with two different oscillation periods. The resonance between the ALP field and noble-gas atoms causes a significant oscillation of the ``noble-gas pendulum". Subsequently, due to the Fermi-contact coupling between the noble gas and the alkali metal atoms (shown as a spring), the alkali metal atoms are pulled to oscillate.

The \text{Rb} spins and noble gas $^{21}\text{Ne}$ spins both react to the oscillating usual magnetic field and generate their own responses. Unlike the case in the self-compensation comagnetometer, where the resonance frequencies of noble-gas atoms and alkali atoms are close to each other in the near-zero field condition, here, the bias field is large enough that noble-gas atoms and alkali atoms are decoupled. In the \(^{21}\)Ne-Rb-K experiment, due to the long coherence time of the \(^{21}\)Ne nuclear spins (\(T_\text{2n} \sim \)40\,s), the \(^{21}\)Ne spins have a sharp resonance line and therefore can be considered as having a discrete system. The Rb-K spins have a shorter coherence time (\(T_\text{2e} \sim \)1\,ms), and thus can be considered as having a continuous spectrum. Assuming the apparatus is sensitive in $\hat{\mathbf{y}}$ direction, we have the result of output effective magnetic field 

\begin{align}
B^\text{eff}_{\rm y}&=\frac{1}{2}\gamma_\text{Ne}B_0^{\rm n} B_{\rm y}\frac{R_2^{\rm n}\sin(\omega t)-(\omega-\omega_0)\cos(\omega t)}{(R_2^{\rm n})^2+(\omega-\omega_0)^2}\notag\\
&+B_{\rm y} \cos(\omega t)\,,
\label{eq:By_effective_field}
\end{align}

\noindent with $B_{\rm y}$ being the measured magnetic field. Therefore the profile of amplification reads 
\begin{align}
\eta_{\rm F}(f)\equiv\frac{\left|B^\text{eff}_{\rm y}\right|}{\left|B_{\rm y}\right|}=\sqrt{\frac{\Gamma^2+(q\Gamma+f_0-f)^2}{\Gamma^2+(f-f_0)^2}}\,,
\label{eq:fano_amplification}
\end{align}

\noindent where $f_0$ is the resonant frequency and the relations $R_2^{\rm n}=(T_\text{2n})^{-1}=2\pi \Gamma$ and amplification factor $q=\frac{1}{2}\gamma _\text{Ne}B_0^{\rm n}T_\text{2n} $ have been used. 

ALP field $\beta_{\rm y}^{\rm{Ne}}$ couples only to noble-gas nuclear spins. The resulting effective magnetic field $\beta_{\rm y}^{\rm{eff}}$ is similar to Eq.\,\eqref{eq:By_effective_field} without the $\beta_{\rm y}^{\rm{Ne}} \cos{\omega t}$ term. The amplification of the ALP field reads 

\begin{align}
\eta_{\rm F}(f)\equiv\frac{\left|\beta^\text{eff}_{\rm y}\right|}{\left|\beta_{\rm y}^\text{Ne}\right|}=\sqrt {\frac{{{ q^2 \Gamma^2}}}{{{\Gamma ^2} + {{\left( {f - {f_0}} \right)}^2}}}},
\label{eq:ALP amplification}
\end{align}

\noindent which is the same as the Lorentzian lineshape.

One can easily see that in the large-$q$ limit the amplification profile reduces to the Lorentzian shape. Using Bloch equations we can obtain the responses to usual magnetic field and ALP field coupling to noble-gas nuclear spins. We use usual magnetic field to calibrate the response of the system. The measured response to usual magnetic field is fitted with Eq.~\eqref{eq:fano_amplification}. And then we substitute the measured parameters into the response equation for ALP field coupling to noble-gas nuclear spins based on Eq.~\eqref{eq:ALP amplification}.

\vspace{4mm}

\noindent \textbf{Stochastic Properties of the Signal}.
In this section, we implement the gradient axion stochastic spectrum derived from Ref.\,\cite{Lee:2022vvb}, while making minor adjustments to account for the new response. In our NMR setup, the comagnetometer is simultaneously sensitive to the anomalous magnetic fields along both $\hat{x}$ and $\hat{y}$ directions, which is different from Ref.\,\cite{Lee:2022vvb} with only one sensitive axis. Therefore, the calculation details need to be adapted accordingly.

The gradient of the axion field within the Galactic halo is assumed to be 
\begin{align}
& \boldsymbol{\nabla}a(t) = \sum_{\Omega_\mathbf{p}} \frac{\sqrt{\rho_\textrm{DM} f(\mathbf{p}) (\Delta p)^3}}{\omega_\mathbf{p}}
\alpha_\mathbf{p} \cos(p^0 t + \Phi_\mathbf{p}) \mathbf{p}  ,\\
&\approx \sum_{\Omega_\mathbf{p}} \sqrt{\rho_\textrm{DM} f(\mathbf{v}) (\Delta v)^3} \, \alpha_\mathbf{v} \times \cos\left(m_a \left(1 + \frac{1}{2}v^2\right)t + \Phi_\mathbf{v}\right)\mathbf{v}, \nonumber
\end{align}
where $\rho_{\rm DM}=0.4~\text{GeV}\cdot\text{cm}^{-3}$~\cite{deSalas:2020hbh} represents the local DM density. Here, $\omega_\mathbf{p}$ denotes the axion energy associated with momentum $\mathbf{p}$, $\alpha_\mathbf{p}$ follows standard Rayleigh distribution as a random variable, and $\Phi_\mathbf{p}$ is a random phase uniformly distributed within $\left[0,2\pi\right]$. The summation runs over all the infinitesimal momentum space $\Omega_\mathbf{p}$ domain.

The DM Maxwell-Boltzmann distribution in the Galactic coordinate system is expressed as~\cite{Lee:2022vvb}
\begin{align}
f(\mathbf{v})\,d^3v =
\dfrac{1}{(2\pi\sigma_\textrm{v}^2)^{3/2}}\exp\left(-\dfrac{(\mathbf{v} + \mathbf{v}_E)^2}{2\sigma_\textrm{v}^2}\right)\, d^3v \,,
\label{eq:mb_velocity_dist}
\end{align} 
where $\mathbf{v}_E$ is the Earth velocity and $\mathbf{v}$ represent the  DM velocity in the lab frame, thus $\mathbf{v}_E+\mathbf{v}$ corresponds to the axion DM velocity in the rest frame of the Galaxy. The parameter $\sqrt{2}\sigma_{\rm v}\approx ~220\, \text{km}/\text{s}$ represents the virial velocity.

By choosing the Cartesian coordinate system $\left\{\hat{u},~\hat{\mathscr{\nu}},~\hat{s} \right\} $, where $\hat{\nu}$ is parallel to $\mathbf{v}_E$ and $\hat{u}$ and $\hat{s}$ form the other orthonormal basis vectors, and applying certain mathematical transformations, we can derive the expression as follows:
\begin{align}
\left(\boldsymbol{\nabla}a\right)^i(t)
&= \sum_j \sqrt{\pi \rho_\textrm{DM} f_j \Delta v}~v_j^2 \epsilon_{i, j} \alpha_{i, j} \cos(\omega_j t + \phi_{i, j}),
\label{eq:axion_grad_radial}
\end{align}
where $j$ is the velocity index and $f_j\equiv (2\pi \sigma_{\rm v}^2)^{-3/2}\exp\left[-\frac{v_j^2+v_E^2}{2\sigma_{\rm v}^2}\right]$. The detailed derivation and the complex definition of $\epsilon_{i,j}$ are provided in Ref.\,\cite{Lee:2022vvb}.

A bias magnetic field is applied along the laboratory $\hat{{\rm z}}$ axis (outward from the ground), and the response signal of the alkali metal ensemble is detected along the $\hat{{\rm x}}$ axis (east-westward). The response of the alkali atom $P_x^e$ receives contributions from the gradient axion field along the $\hat{{\rm x}}-\hat{{\rm y}}$ plane.
Based on the argument below Eq.\,\eqref{eq:pi/2 argue}, there exists a $\pi/2$ phase difference between the responses $\mathbf{b}_0\cdot\hat{{\rm x}}$ and $\mathbf{b}_0\cdot\hat{{\rm y}}$, which need to be addressed in the calculations of ALP signal response variance.

For the case of ALP DM, we have $b_0^{\rm x}=\vec{\nabla} a\cdot \mathbf{\hat{m}}_{\rm x}$ and $b_0^{\rm y}=\vec{\nabla} a\cdot \mathbf{\hat{m}}_{\rm y}$. The sensitive axes $\mathbf{\hat{m}}_{\rm x}$ and $\mathbf{\hat{m}}_{\rm y}$ in the Cartesian coordinate system $\left\{u,~v,~s \right\}$ during the experiment can be described as follows:
\begin{align}
\hat{\mathbf{m}}_i^{\rm x}(t)&= C_i^{(1)} \cos(\omega_{\rm e} t+\theta_i^{(1)}) + D_i^{(1)},
\nonumber\\
\hat{\mathbf{m}}_i^{\rm y}(t)&= C_i^{(2)} \cos(\omega_{\rm e} t+\theta_i^{(2)}) + D_i^{(2)},
\label{eq:axis_approx}
\end{align}
where $i\in \left\{u,~v,~s\right\}$, $\omega_{\rm E}= 2\pi \times 1.16~\mu\text{Hz}$ is the Earth's sidereal frequency, and the coefficients $C_i^{(1)}, C_i^{(2)}, D_i^{(1)}, D_i^{(2)}, \theta_i^{(1)},$ and $\theta_i^{(2)}$ are determined through coordinate transformations from the laboratory coordinate system to the celestial coordinate system, and subsequently to the Galactic coordinate system. These coefficients are calculated before each individual scanning.

\vspace{4mm}

\noindent \textbf{The Calculation for the Axion-Nucleon Coupling Sensitivity}.
\label{sec:calculation}

We conduct a profile analysis using the Log-Likelihood Ratio (LLR) method based on the response power spectrum $(V^2)$. In this experiment, we are in the peaked regime, indicating that the power spectrum exhibits a peak at the resonant frequency even without injecting a magnetic field in the $x-y$ plane. This phenomenon arises because the amplified magnetic noise after amplification far surpasses the photon shot noise, which cannot be amplified. Therefore, we cannot assume that the background noise is frequency-independent white noise within the narrow LLR frequency range (as assumed in Refs.\,\cite{Jiang:2021dby, Bloch:2021vnn, Lee:2022vvb, Wei:2023rzs}). Instead, we model the background noise as a sum of two zero-mean Gaussian components: one being the non-amplifiable photon shot noise with a variance of $\Sigma_{\rm na}$, which remains constant in frequency, and the other being the amplifiable magnetic noise with a variance of $\Sigma_{\rm a} \cdot\eta_{\rm F}(f)^2$, following the Fano profile. The axion signal is modeled as a multivariate Gaussian distribution~\cite{Bloch:2021vnn,Lee:2022vvb}, with a variance of $\Sigma_{\rm A}(f_k,f_r)=g_\text{cali}(f_k)g_\text{cali}(f_r)\mathbf{\sigma}_{\rm A}(f_k,f_r)$. 

The calibration function $g_\text{cali}(f)$ is the transfer function that converts magnetic field strength in Tesla to the apparatus response in Voltage at a given frequency $f$, which includes the amplification function for the signal. We obtain this function by injecting magnetic fields of known amplitude and varying frequency along the sensitive axes, subsequently measuring the apparatus response. We fit the obtained response using a Fano lineshape given by Eq.\,\eqref{eq:fano_amplification}. From the fit parameters $q$ and $\Gamma$, we can then derive the calibration function for the ALP field, which is modeled by a Lorentzian lineshape. The complete response of the ALP field is depicted in Fig.\,\ref{fig:response to B and bn}. For a narrow band near the resonant frequency (e.g., $f_0 \pm 0.25$ Hz), the shape is accurately described by the Lorentzian function. Lastly, the aforementioned calibration procedure is conducted before each measurement.

The covariance matrix $\sigma_{\rm A}(f_k,f_r)$ is defined by Eqs. B33 and B34 in Ref.~\cite{Lee:2022vvb}:
 \begin{align}
\nonumber
\sigma(A_k, A_r) &= 4\pi\rho_\textrm{DM} \left(\frac{g_\textrm{aNN}}{T\mu_\textrm{Ne}}\right)^2 \sum_i \int_0^\infty dv \, f(v) \epsilon_i^2(v) \\
& \hspace{0.2cm} \times v^4\left[E_{ik}(v)E_{ir}(v) + F_{ik}(v)F_{ir}(v)\right],
\label{eq:covAA_def}
\end{align}
\begin{align}
\nonumber
\sigma(A_k, B_r) &= 4\pi\rho_\textrm{DM} \left(\frac{g_\textrm{aNN}}{T\mu_\textrm{Ne}}\right)^2 \sum_i \int_0^\infty dv \, f(v) \epsilon_i^2(v) \\
& \hspace{0.2cm} \times v^4 \left[E_{ik}(v)F_{ir}(v) - F_{ik}(v)E_{ir}(v)\right],
\label{eq:covAB_def}
\end{align}
\begin{align}
\nonumber
\sigma(B_k, A_r) &= -\sigma(A_k, B_r)=\sigma(A_r, B_k),\\
\sigma(B_k, B_r) &= \sigma(A_k, A_r).\nonumber
\end{align}
where $\mu_{\rm Ne}=2\pi\times 3.36\text{MHz}\cdot\text{T}^{-1}$ is the $^{21}\text{Ne}$ gyromagnetic ratio and
\begin{equation}
    A_k = \frac{2}{N}\Re[\tilde{\beta}_k] \,,\, B_k = -\frac{2}{N}\Im[\tilde{\beta}_k]\,,
    \label{eq:Ak_Bk_def}
\end{equation}
and $\tilde{\beta}_k$ is the discrete Fourier transform of obtained data time series $\beta_{n}$ 
\begin{equation}
    \tilde{\beta}_k = \sum_{n=0}^{N-1} \beta_{n} \exp^{-{\rm i}\omega_k n \Delta t} \,,\, \omega_k \equiv \frac{2{\rm \pi} k}{N\Delta t}\,.
    \label{eq:beta_fft_def}
\end{equation}
The definitions of $E_{ik}(v)$ and $F_{ik}(v)$ are slightly modified to account for two sensitive axes and their ${\rm \pi}/2$ relative phase in the response of $P^{\rm e}_{\rm x}$:

\begin{align}
& E_{i,k}(v) = \nonumber\\
\nonumber
& \frac{\sin{\left(\frac{T \omega_{\rm E}}{2} + \frac{T \omega}{2} - \frac{T \omega_{k}}{2} \right)}}{2 \left(\omega_{\rm E} + \omega - \omega_{k}\right)}\left[ {C}_i^{(1)}\cos{\left(\frac{T \omega_{\rm E}}{2} + \frac{T \omega}{2} - \frac{T \omega_{k}}{2} + {\theta}_{i}^{(1)} \right)}\right.\\
&\left.-{C}_i^{(2)}\sin{{\left(\frac{T \omega_{\rm E}}{2} + \frac{T \omega}{2} - \frac{T \omega_{k}}{2} + {\theta}_{i}^{(2)} \right)}}\right] \nonumber\\
& + \frac{\sin{\left(\frac{T \omega}{2} - \frac{T \omega_{k}}{2} \right)}} {\omega - \omega_{k}}\left[ {D}_{i}^{(1)}\cos{\left(\frac{T \omega}{2} - \frac{T \omega_{k}}{2} \right)}-\right.\nonumber\\
&\left. {D}_{i}^{(2)}\sin{\left(\frac{T \omega}{2} - \frac{T \omega_{k}}{2} \right)}\right]-\nonumber\\
& \frac{\sin{\left(\frac{T \omega_{\rm E}}{2} - \frac{T \omega}{2} + \frac{T \omega_{k}}{2} \right)} }{2 \left(- \omega_{\rm E} + \omega - \omega_{k}\right)}\left[{C}_{i}^{(1)}\cos{\left(\frac{T \omega_{\rm E}}{2} - \frac{T \omega}{2} + \frac{T \omega_{k}}{2} + {\theta}_{i}^{(1)} \right)}\right.\nonumber\\
&\left.+{C}_{i}^{(2)}\sin{\left(\frac{T \omega_{\rm E}}{2} - \frac{T \omega}{2} + \frac{T \omega_{k}}{2} + {\theta}_{i}^{(2)} \right)} \right],
\label{eq:E_ijk_def}
\end{align}

\begin{align}
& F_{i,k}(v) = \nonumber\\
& \frac{\sin{\left(\frac{T \omega_{\rm E}}{2} + \frac{T \omega}{2} - \frac{T \omega_{k}}{2} \right)}}{2 \left(\omega_{\rm E} + \omega - \omega_{k}\right)}\left[ -{C}_i^{(1)}\sin{\left(\frac{T \omega_{\rm E}}{2} + \frac{T \omega}{2} - \frac{T \omega_{k}}{2} + {\theta}_{i}^{(1)} \right)}\right.\nonumber\\
&\left.-{C}_i^{(2)}\cos{{\left(\frac{T \omega_{\rm E}}{2} + \frac{T \omega}{2} - \frac{T \omega_{k}}{2} + {\theta}_{i}^{(2)} \right)}}\right] \nonumber\\
\nonumber
& + \frac{\sin{\left(\frac{T \omega}{2} - \frac{T \omega_{k}}{2} \right)}} {\omega - \omega_{k}}\left[ -{D}_{i}^{(1)}\sin{\left(\frac{T \omega}{2} - \frac{T \omega_{k}}{2} \right)}-\right.\nonumber\\
&\left. {D}_{i}^{(2)}\cos{\left(\frac{T \omega}{2} - \frac{T \omega_{k}}{2} \right)}\right]+\nonumber\\
& \frac{\sin{\left(\frac{T \omega_{\rm E}}{2} - \frac{T \omega}{2} + \frac{T \omega_{k}}{2} \right)} }{2 \left(- \omega_{\rm E} + \omega - \omega_{k}\right)}\left[-{C}_{i}^{(1)}\sin{\left(\frac{T \omega_{\rm E}}{2} - \frac{T \omega}{2} + \frac{T \omega_{k}}{2} + {\theta}_{i}^{(1)} \right)}\right.\nonumber\\
&\left.+{C}_{i}^{(2)}\cos{\left(\frac{T \omega_{\rm E}}{2} - \frac{T \omega}{2} + \frac{T \omega_{k}}{2} + {\theta}_{i}^{(2)} \right)} \right]
\label{eq:F_ijk_def}
\end{align}
with $\omega=m_a(1+\frac{1}{2}v^2)$.

Since both the signal and the background are multivariate Gaussian random variables, we can construct the likelihood function as follows: 
        \begin{align}
            L(\textbf{d}|g_{\rm aNN}, \Sigma_{\rm na},\Sigma_{\rm a})=\frac{1}{\sqrt{(2\pi)^{2N}\mathrm{det}(\Sigma)}} \mathrm{exp}\left(-\frac{1}{2}\textbf{d}^T\Sigma^{-1}\textbf{d}\right).
        \end{align}
with vector ${\bf d} = \left\{ A_k, B_k\right\} $ and variance matrix $\Sigma=\Sigma_{\rm A}(g_{\rm aNN})+\Sigma_{\rm na}\cdot \mathbbm{1}+\text{diag}(\Sigma_{\rm a} \cdot \eta_{\rm F}^2(f))$

To set a quantitative limit, we use the LLR defined below
                \begin{align}
                    {\rm LLR}(g_{\rm aNN}) =
                    \begin{cases}
                        -2\log\left[\frac{L(g_{\rm aNN},\tilde{\Sigma}_{\rm a},\tilde{\Sigma}_{\rm na})}{L(\hat{g}, \hat{\Sigma}_{\rm a},\hat{\Sigma}_{\rm na})}\right],&\hat{g}\leq  g_{\rm aNN} \\
                        0, &\hat{g}> g_{\rm aNN}
                    \end{cases}
                \end{align}
where we treat both $\Sigma_{\rm a}$ and $\Sigma_{\rm na}$ as nuisance parameters. $\tilde{\Sigma}_{\rm a}$ and $\tilde{\Sigma}_{\rm na}$ maximizes the likelihood $L$ for a fixed nucleus coupling $g_{\rm aNN}$ and $\hat{\Sigma}_{\rm a}$, $\hat{\Sigma}_{\rm na}$ and $\hat{g}$ are unconditional estimators that maximize likelihood $L$ for given $\bf d$. In order to establish an upper limit, we set ${\rm LLR}=0$ if $\hat{g} > g_{\rm aNN}$. The defined $ {\rm LLR}$ can be asymptotically shown to follow half-$\chi^2$ distribution with the cumulative distribution function~\cite{Cowan:2010js}
\begin{align}
    \Phi( {\rm LLR}(g_{\rm aNN})\le y)=\frac{1}{2}\left[1+\text{erf}\left(\sqrt{\frac{y}{2}}\right)\right] ,
    \label{eq:cdf}
\end{align}
thus the $95\%$ CL upper limits are determined by finding the value of $g_{\rm aNN}$ where ${\rm LLR}(g_{\rm aNN})=2.7055$.
The effective coupling between axions and the nucleus $N$ is given by $g_{\rm aNN} = \xi_{\rm n} g_{\rm ann} + \xi_{\rm p} g_{\rm app}$, where $\xi_{\rm n}^{\rm Ne} = 0.58$ and $\xi_{\rm p}^{\rm Ne} = 0.04$ are the spin-polarization fractions for neutrons and protons in ${}^{21}{\rm Ne}$~\cite{Brown:2016ipd, Almasi:2018cob}. When setting limits on the coupling between ALP and neutrons, we set $g_{\rm app}=0$. Similarly, when setting the limits on the coupling of protons, we set $g_{\rm ann}=0$.

\vspace{4mm}

\noindent \textbf{The Shape of the Sensitivity Curve}.
For measuring times smaller or equal to the coherent time $\tau_a \sim 2 \hbar/(m_a v_{\rm vir}^2)\sim 165~ \text{hours} \left(\frac{\text{Hz}}{f_a}\right)$, the signal predominantly falls within a single bin. As discussed in the main text, in this scenario, the lineshape of the signal limit can be estimated.

In each single bin, both the signal and the background are treated as the Gaussian variables in the frequency domain.
To establish the upper limit for the signal strength, we make the assumption that the data consists only background with a certain variance.
Therefore, the data power spectrum follows a $\chi^2$ distribution with two degrees of freedom because the variance is complex. The corresponding cumulative distribution function follows an exponential distribution~\cite{Walck:1996cca},
\begin{align}
    P({\rm Power}(f)\le y)=1-{\rm e}^{-y/2\sigma^2}.
\end{align}

\begin{figure}[tbh]
    \centering
    \includegraphics[width=0.99\columnwidth]{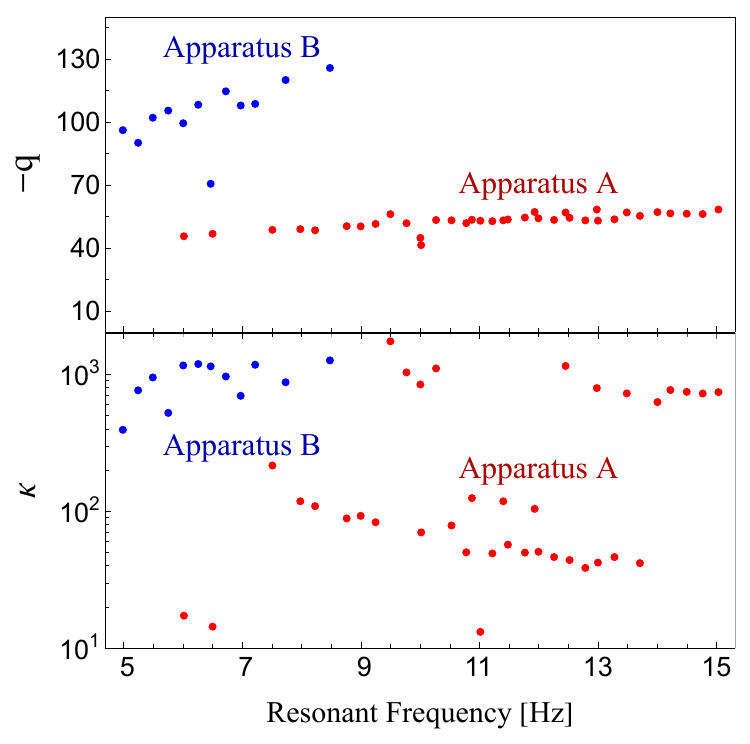}
    \caption{\textbf{
    The amplification factor $q$ and amplification level $\kappa$.} $q$ and $\kappa$ were determined for the 49 individual scans, with red and blue dots representing apparatus A and B, respectively. 
    The values of $q$ and $\Gamma$ were obtained from the calibration process, and the value of $\kappa$ was extracted from each experiment data.
    Apparatus A exhibited $\left| q \right| \sim 40-50$ and $\kappa$ ranging between 10--1000, while apparatus B had $\left|q\right| \sim 100$ and $\kappa$ ranging between 400--1000. 
    }
    \label{fig:amplification_all}
\end{figure}

Therefore, the $95\%$ confidence level limit on the variance contributed by the signal is $y_{95\%} = 5.99~\sigma^2$. Consequently, the signal coupling at the frequency bin $f$ should satisfy:
\begin{align}
5.99~\left[\Sigma_{\rm a}\eta_{\rm F}(f)^2+\Sigma_{\rm na}\right]=c_1\eta(f)^2\left(g_{\rm aNN}^{95\%}(f)\right)^2,
    \label{eq:threshold}
\end{align}
$\Sigma_{\rm a}$ and $\Sigma_{\rm na}$ denote the variances of the amplifiable and non-amplifiable background components, e.g. the magnetic noise and the photon shot noise respectively. $c_1$ represents the signal coefficient independent of $f$. $\eta_{\rm F}(f)$ and $\eta(f)$ correspond to the Fano profile and Lorentzian profiles, respectively. In this paper, we assume that the axion DM-electron coupling is negligible, resulting in no interference between alkali metal and noble gas responses. Therefore, the signal profile can be described by a Lorentzian shape at the leading order, which is symmetric near the resonant frequency. As illustrated in Fig.\,\ref{fig:response to B and bn}, the response to the ALP field does show a Lorentzian shape and only deviates in the region far from the resonance, which is different from the usual magnetic field. Since we are primarily interested in a narrow band near the resonant frequency, e.g., $f_0 \pm 0.25$ Hz, the Lorentzian profile is a good description for the signal response.
Following Eq.\,\eqref{eq:threshold}, we can derive the sensitivity as a function of frequency $g_{\rm aNN}(f)$ and its FWHM for the signal as 
 \begin{align}
     &g_{\rm aNN}^{95\%}(f)= \sqrt{\frac{5.99\Sigma_{\rm a}}{c_1} }  
     \sqrt{1+\frac{1+\Delta^2}{\kappa}+\frac{1+\Delta^2 - 2 q \Delta}{q^2} },
     \label{eq:real-gaNN}
     \\
     & \Gamma_{\rm FWHM}=2\sqrt{3}\Gamma 
     \sqrt{1+\kappa \left(1 - 
     \frac{\kappa (\kappa + 2 q^2)}{(\kappa + q^2)^2} \right) } ,
     \label{eq:real-FWHM}
 \end{align}
where $\Delta \equiv (f-f_0)/\Gamma$, $q$ represents the amplification factor at the Larmor frequency, and $\kappa\equiv q^2\Sigma_{\rm a}/\Sigma_{\rm na}$ is the amplification level.

In practice, we obtain values for $q$ and $\Gamma$ from the calibration process, and $\kappa$ from the experimental data. In Fig.\,\ref{fig:amplification_all}, we present the amplification factor $q$ and the amplification level $\kappa$ for the 49 individual scans, represented by red and blue dots for apparatus A and B, respectively. Apparatus A has an amplification factor of approximately $- q\approx 50$, while Apparatus B has a value of around 100. The amplification factor $q$ is negative because the dip, attributed to the Fano effect, appears on the left side of the resonant frequency, see Fig.\,\ref{fig:response to B and bn}.
As for the amplification level $\kappa$, Apparatus B shows a range between 400--1000, whereas Apparatus A mostly falls within the range of 100 and 1000. 
Since we have observed in the experiment that the photon shot noise is much larger than the magnetic noise ($\Sigma_{\rm na} \gg \Sigma_{\rm a}$), we can make the approximation $q^2 \gg \kappa$, as  supported by Fig.\,\ref{fig:amplification_all}. Using this approximation, the sensitivity $g_{\rm aNN}^{95\%}(f)$ and its bandwidth $\Gamma_{\rm FWHM}$ in Eqs.\,\eqref{eq:real-gaNN} and \eqref{eq:real-FWHM},  will revert to the simple Lorentzian expression.

Due to the Fano effect, there is another difference from the simple Lorentzian case: the best sensitivity for $g_{\rm aNN}^{95\%}(f)$ is not achieved at the resonant frequency $f_0$. Following Eq.~\eqref{eq:real-FWHM}, we can determine that the best sensitivity $\tilde{g}_{\rm aNN}^{\rm min} \equiv g_{\rm aNN}^{95\%}(\tilde{f}_0)$ and its corresponding frequency $\tilde{f}_0$ are
\begin{align}
     \tilde{g}_{\rm aNN}^{\rm min} & = \sqrt{\frac{5.99\Sigma_{\rm a}}{c_1} }  
     \sqrt{1+\kappa^{-1} + \frac{q^2 + \kappa - q^2 \kappa}{q^2 \left(q^2  +  \kappa \right)} } , \nonumber \\
     & \approx  \sqrt{\frac{5.99\Sigma_{\rm a}}{c_1} }  
     \sqrt{1+\kappa^{-1}  } , \\
     \tilde{f}_0 & = f_0 + \Gamma \frac{\kappa q}{\kappa + q^2} .
\end{align}
In the second line, the best sensitivity $\tilde{g}_{\rm aNN}^{\rm min}$ is very close to the Lorentzian value when $q^2 \gg \kappa$. The frequency $\tilde{f}_0$ deviates from its Lorentzian value $f_0$, which is significant and can be observed.

\begin{figure}[tbh]
    \centering
    \includegraphics[width=0.99\columnwidth]{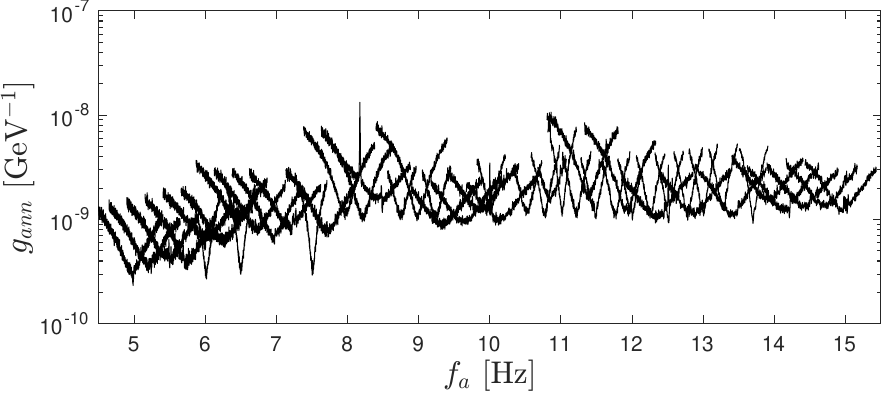}
    \caption{\textbf{The sensitivity curves.} The curves were plotted for a total of 49 individual experimental scans, each smoothed using a $\pm 250$ bins moving average.
    }
    \label{fig:individual_median_result}
\end{figure}

In Fig.\,\ref{fig:individual_median_result}, we present a summary plot of the sensitivity curves for all 49 individual scans, each smoothed with a $\pm 250$ bins moving average. Each sensitivity curve exhibits a peak centered around its resonant frequency, but the peaks vary in width, corresponding to their respective $\kappa$ values. 
Their widths are clearly around 0.25 Hz, which are much broader than the typical nuclear spin resonance width for ${}^{21}$Ne. Specifically, there are four points for Apparatus A with $\kappa \sim 10$, which exactly match the sharp peaks in Fig.\,\ref{fig:individual_median_result}. 
These points at around 6\,Hz, 6.5\,Hz, and 7.5\,Hz are better than the nearby scans, due to the suppression of the experimental environmental noises. 

In addition, our experiment operates within a frequency range of 5-15 Hz. We maintained stable experimental conditions throughout the measurements, allowing us to estimate the measurement error of the transfer function $g_{\rm cali}$ at key frequency points (5 Hz, 10.75 Hz, 11.5 Hz, and 13.75 Hz). Assuming the error trend follows a linear behavior across these points, we extrapolate it to estimate the systematic error for the entire range. 

\begin{figure*}[tbh]
    \centering
    \includegraphics[width=1.5 \columnwidth]{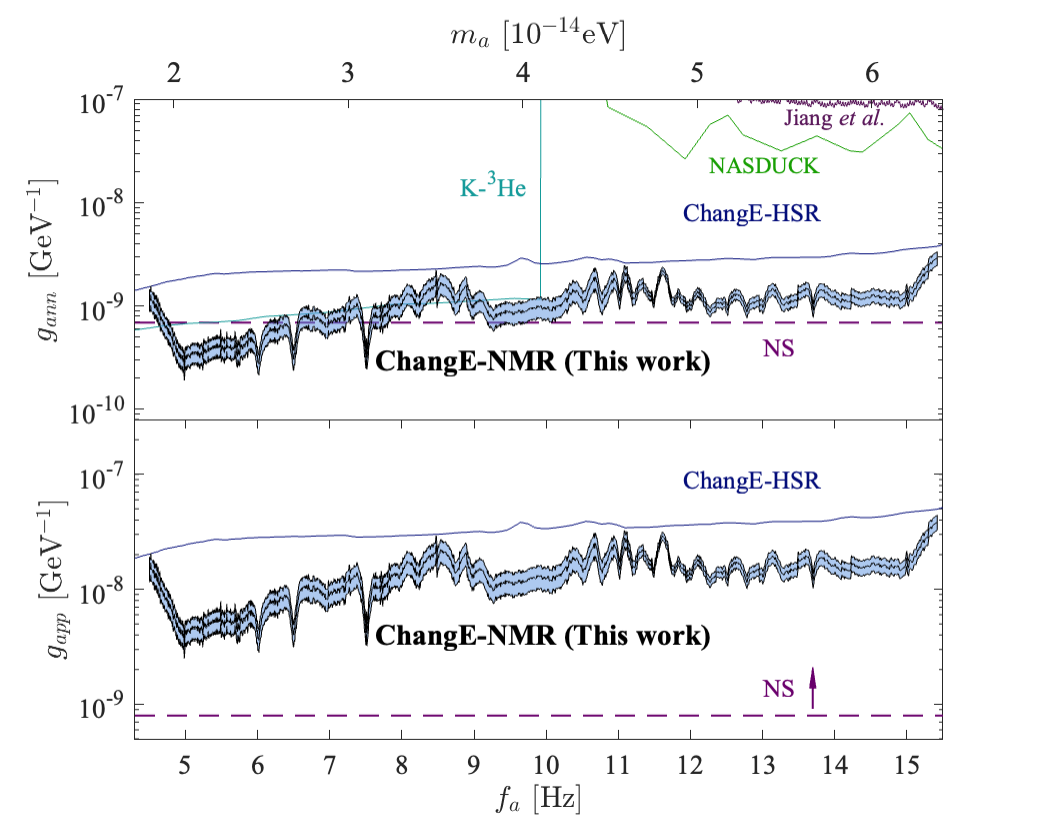}
    \caption{\textbf{
    The systematic uncertainty of sensitivity from $g_{\rm cali}$.} 
The blue bands in the figure denote the sensitivity standard deviation from systematic uncertainties from the calibration function $g_{\rm cali}$. Since experimental conditions are maintained stable throughout the measurements, we estimate the measurement error of the transfer function $g_{\rm cali}$ at several key frequency points then linearly extrapolate the values to estimate the systematic error for the small frequency range of $5-15$ Hz. The description of other elements are the same as Figure.~\ref{fig:finalsensitivity}. }
    \label{fig:deviation}
\end{figure*}

For each key frequency point, we performed three independent measurements of $g_{\rm cali}$. The systematic error of our measurements is defined as the error of $g_{\rm cali}$, and we use it to calculate the standard deviation of neutron coupling $g_{\rm ann}$. In Fig.~\ref{fig:deviation}, we show the sensitivity with the error bands reflecting systematic uncertainties from the calibration function.

\noindent \textbf{The Scan Strategy}.
As discussed in Refs.~\cite{Lee:2022vvb, Dror:2022xpi, Yuzhe2023}, when the transverse spin-relaxation time is significantly shorter than both the ALP DM coherent time $\tau_a$ and the measurement time $T$ (as is the case in this experiment), the $95\%$ upper limits follow a power law of $g \propto T^{-1/2}$ and $T^{-1/4}$ for $T \ll \tau_a$ and $T \gg \tau_a$, respectively. Since our experiment operates at low frequencies, approximately on the order of ${\mathcal O}(10)$ Hz, the coherent time can extend up to around $\sim 30$ hours. Therefore, we choose the optimal measurement time as $T \simeq \tau_a$ for each run, with a cumulative measurement duration of 1440 hours for 49 individual scans.

As explained in the previous chapter, given that we are in the magnetic-dominated regime, widening the scan step can be considerably beneficial, potentially leading to a reduction in the number of scans needed. In practice, we opt for a frequency step of 0.25 Hz to scan the ALP frequency range between $5-15$ Hz.

\vspace{4mm}

\noindent \textbf{The Post-Analysis of Possible Candidates}.
Since the axion mass is not known as a prior, one has to test all the frequencies in a given range. We explore if there are possible axion DM signal in the data that exclude the background-only hypothesis. For a quantitative analysis of possible axion candidates, we define the following test statistics to test the significance of a best-fit signal compared to the background only model \cite{Lee:2022vvb},
    \begin{align}
        q_{\rm D} =
            \begin{cases}
            -2\log\left[\frac{L(0,\tilde{\Sigma}_{\rm a},\tilde{\Sigma}_{\rm na})}{L(\hat{g}, \hat{\Sigma}_{\rm a},\hat{\Sigma}_{\rm na})}\right],&\hat{g}\leq  0 \\
            0, &\hat{g}> 0
            \end{cases}
    \end{align}
where $\tilde{\Sigma}_{\rm a}$ and $\tilde{\Sigma}_{\rm na}$ are maximum LLR estimators for $g_{\rm aNN}=0$, while $\hat{g}$, $\hat{\Sigma}_{\rm na}$ and $\hat{\Sigma}_{\rm a}$ maximize the LLP across all parameter spaces.

Since we need to test a large number of ALP candidate frequencies within a given frequency range, we have to consider the look-elsewhere effect. In the case of $N_{\rm F}$ independent tests, the global $p$-value, denoted as $p_{\rm global}$, is given by the formula: 
\begin{align}
    p_{\rm global}=1-(1-p)^{N_{\rm F}}  \,,
\end{align}
where $p$ is the $p$-value of a single measurement. For sufficiently small $p$, we can approximate it as $p_{\rm global}=1-N_{\rm F} p$. Utilizing Eq.\,\eqref{eq:cdf}, we can calculate the corresponding $q_{\rm global}$ after accounting for the look-elsewhere effect:
\begin{align}
    q_{\rm global}=2\left[{\rm erf}^{-1}\left(1-2\frac{p_{\rm global}}{N_{\rm F}} \right)\right]^2 \, .
\end{align}

To determine the number of independent tests, $N_F$, for the frequencies, we employ the method outlined in Ref.\,\cite{Lee:2022vvb}. $N_F$ is given by:
\begin{align}
    N_{\rm F}=\int^{{\rm log~max} (F) }_{{\rm log~min} (F) }\frac{1}{\alpha(v)v_0^2} dv.
\end{align}
where $F$ represents the test frequency interval for ALP. The function $\alpha(v)$ is determined through Monte-Carlo simulations. Using the above equations, the $5\sigma$ threshold corresponds to $q_{\rm global}$ within the range $[41.33,~47.36]$ for 49 individual measurements.

After analyzing all 49 individual scans, several frequency candidates surpass the threshold of the background-only hypothesis, ${\rm LLR}_{\rm discovery} = q_{\rm global}$, signifying a $5\sigma$ significance with consideration for the look-elsewhere effect. However, each of these candidate frequencies can be tested by multiple adjacent individual scans. Upon closer examination, we observed that these $5\sigma$ anomalous peaks were not consistently present in other individual scans, suggesting they may be transient background fluctuations. Therefore, we conclude that there is no ALP signal in our experiments.

\vspace{4mm}

\noindent \textbf{Data availability}

All relevant data are available from the corresponding author upon request.

\noindent \textbf{References}

\bibliography{mainrefs}

\vspace{4mm}

\noindent \textbf{Acknowledgments}

The work of KW is supported by NSFC under Grants No. 62203030 and 61925301 for Distinguished Young Scholars and in part by the Innovation Program for
Quantum Science and Technology under Grant 2021ZD0300401. 
The work of JL is supported by NSFC under Grant No. 12075005, 12235001. 
The work of X.P.W. is supported by the National Science Foundation of China under Grant No. 12005009 and the Fundamental Research Funds for the Central Universities. 
The work of WJ and DB is supported by the DFG Project ID 390831469: EXC 2118 (PRISMA+ Cluster of Excellence), by the COST Action within the project COSMIC WISPers (Grant No. CA21106), and by the QuantERA project
LEMAQUME (DFG Project No. 500314265).

\vspace{4mm}

\noindent \textbf{Author contributions}

Kai Wei, Wei Ji, Jia Liu, Xiaoping Wang and Dmitry Budker proposed the experimental concept and devised the experimental protocols, and wrote the manuscript. Zitong Xu, Kai Wei, Xing Heng and Xiaofei Huang designed and performed experiments, analyzed the data, and wrote the manuscript. Xiaolin Ma, Yuxuan He, Tengyu Ai, Jian Liao, and Jia Liu analyzed the data, added theoretical discussion, and edited the manuscript. All authors contributed with discussions and checking the manuscript. 

\vspace{4mm}

\noindent \textbf{Competing interests} 

The authors declare that they have no competing interests.

\end{document}